\documentclass[sn-mathphys]{sn-jnl}% Math and Physical Sciences Reference Style
\jyear{2021}%

\theoremstyle{thmstyleone}%
\theoremstyle{thmstyletwo}%

\theoremstyle{thmstylethree}%

\def\vii{{\bm v}}

\def\BBb{{ B}}

\def\ace{\varphi}

\def\cBB{c}

\newcommand{\yan}[1]{{\footnotesize \color{purple} [Yan: #1]}}

\begin{document}

\title[Atomic cluster expansion for Pt-Rh catalysts]{Atomic cluster expansion for Pt-Rh catalysts: From ab initio to the simulation of nanoclusters in few steps}

%%%%-------------------authors--------------------

\author[1]{\fnm{Yanyan} \sur{Liang}}\email{yanyan.liang@rub.de}

\author[1]{\fnm{Matous} \sur{Mrovec}}\email{matous.mrovec@rub.de}

\author[1]{\fnm{Yury} \sur{Lysogorskiy}}\email{yury.lysogorskiy@rub.de}

\author[2]{\fnm{Miquel} \sur{Vega-Paredes}}\email{m.vega@mpie.de}

\author[2]{\fnm{Christina} \sur{Scheu}}\email{scheu@mpie.de}

\author*[1]{\fnm{Ralf} \sur{Drautz}}\email{ralf.drautz@rub.de}

\affil*[1]{\orgdiv{ICAMS}, \orgname{Ruhr-Universit{\"a}t Bochum}, \orgaddress{\street{Universit{\"a}tsstra{\ss}e 150}, \city{Bochum}, \postcode{44801}, \country{Germany}}}

\affil[2]{\orgname{Max Planck Institut für Eisenforschung}, \orgaddress{\street{Max Planck Stra{\ss}e 1}, \city{Düsseldorf}, \postcode{40237}, \country{Germany}}}


\abstract{
Insight into structural and thermodynamic properties of nanoparticles is crucial for designing optimal catalysts with enhanced activity and stability. We present a semi-automated workflow for parameterizing the atomic cluster expansion (ACE) from ab initio data. The main steps of the workflow are the generation of training data from accurate electronic structure calculations, an efficient fitting procedure supported by active learning and uncertainty indication, and a thorough validation. We apply the workflow to the simulation of binary Pt-Rh nanoparticles that are important for catalytic applications. We demonstrate that the Pt-Rh ACE is able to reproduce accurately a broad range of fundamental properties of the elemental metals as well as their compounds while retaining an outstanding computational efficiency. This enables a direct comparison of simulations to high resolution experiments.  
}

\keywords{atomic cluster expansion, machine learning, atomistic simulations, platinum-rhodium catalyst, nanoclusters}

\maketitle

%%%%%%%%%%%%%%%%%%%%%%%%%%%%%%%%%%%%%%%%%%%%%%%%%%%%%%%%%%%%%%%%%%%%%%%%%%%%%%%%%%%%%%%%%%%%%%%%%%%%%%%%%%%%%%%

\section{Introduction}\label{sec:intro}

In the field of catalysis, noble metals play a key role. Among them, platinum (Pt) and rhodium (Rh) are used extensively as heterogeneous catalysts due to their superior activity and stability~\cite{li2011characterization}. For example, Pt-Rh based materials are widely used as effective vehicle exhaust catalysts to reduce emissions of harmful pollutants CO and NO$_\text{x}$ into less harmful CO$_2$ and N$_2$~\cite{eley1998advances,ravindra2004platinum}. Furthermore, they have also been employed in fuel cells~\cite{jung2014pt,ramli2018platinum} to enhance the oxygen reduction and hydrogen oxidation reactions. 

These catalytic materials are typically applied in the form of nanoparticles or nanoclusters to maximize the active surface area where the chemical reactions take place~\cite{sun2021advancements}.  An important aspect of the nanoparticles is their microstructure that can be tailored to optimize functional properties as well as stability. In multicomponent systems, the goal is often not to achieve a random distribution of chemical elements as in a solid solution, but to design nanoparticles with specific morphologies~\cite{ghosh2012core,gawande2015core}. For instance, despite the fact that Pt and Rh are miscible elements with face-centered cubic (fcc) structures their nanoparticles have usually been prepared in the form of concentric composites, typically with a spherical Rh core surrounded by a Pt-rich shell layer. Such core-shell morphologies can be prepared using a wet chemical approach, where the core is synthesized first and the surface is coated afterwards~\cite{ghosh2012core,gawande2015core}. They can exhibit an enhanced catalytic activity and structural stability compared to traditional alloyed catalysts~\cite{ghosh2012core} as a result of their non-equilibrium crystal structures~\cite{garzon2018controlling} or shapes. However, the non-equilibrium nanoparticles may not be stable under operating conditions and undergo transformations that lead to deterioration of the catalytic activity. 

\begin{figure}
    \centering
    \subfigure[]{
    \includegraphics[scale=0.4]
    {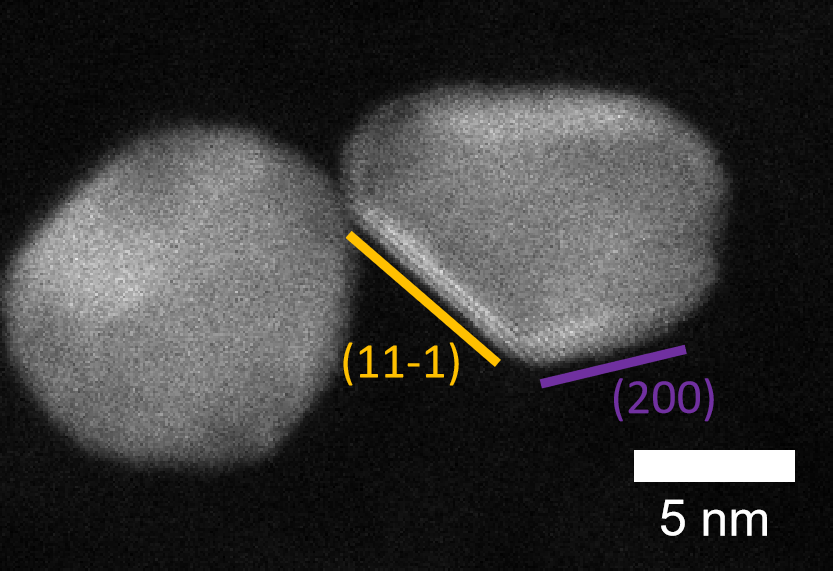}}
    \subfigure[]{
     \includegraphics[scale=0.4]
     {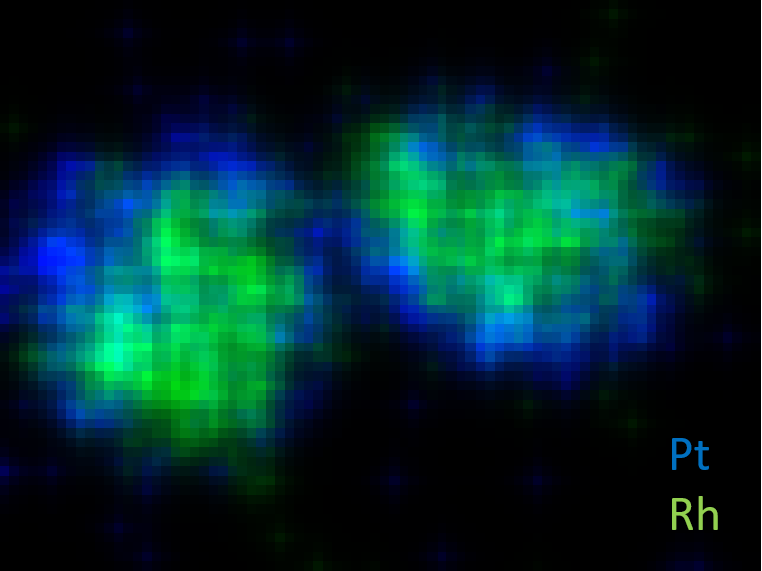}}
    \caption{STEM-HAADF micrograph (a) and corresponding EDS composition map (b) of a Rh-Pt core shell nanoparticle. Pt is in blue and Rh in green. In (a) the crystallographic facets are indicated.}
    \label{fig:Expnano}
\end{figure}

In Fig.~\ref{fig:Expnano}, an example of a Pt-Rh core-shell nanoparticle with a distorted octahedral shape is shown~\cite{vega2023}. The high angle annular dark field (HAADF) micrograph and energy dispersive X-ray spectroscopy (EDS) map acquired in a scanning transmission electron microscope (STEM) reveal a varying Pt shell thickness of 3-5 monolayers. STEM is widely used for characterizing catalyst nanoparticles, since it can provide information on the atomic arrangement and composition~\cite{su2017advanced}. However, investigations of dynamic processes at elevated temperatures requires to carry out challenging and time consuming in-situ experiments that are often limited by beam-induced artefacts~\cite{manjon2022exploring}.  Therefore, it is desirable to complement the experimental characterization with atomistic modeling to obtain a deeper understanding of the thermodynamic and kinetic phenomena governing the structure of nanoparticles.

Atomistic simulations can provide valuable information about energetics of particular structural features, such as formation energies of point defects or various surface facets, but also directly follow the dynamical evolution of the entire system using molecular dynamics (MD) or explore the thermodynamic stability using Monte Carlo (MC) methods. Even though it is nowadays possible to study systems of millions of atoms, approaching realistic dimensions of nanoparticles, the essential prerequisite for reliable quantitative predictions is an accurate and efficient description of interatomic interactions. However, there exist, to our best knowledge, no interatomic potentials for the Pt-Rh binary system that could be applied in large-scale MD or MC studies of thermodynamic and kinetic phenomena. 

In this work, we developed an atomic cluster expansion (ACE) parametrization for Pt-Rh using a semi-automatic workflow consisting of the generation of a training dataset based on accurate density functional theory (DFT)~\cite{hohenberg1964inhomogeneous,kohn1965self} calculations, efficient fitting and validation procedures~\cite{lysogorskiy2021performant}, and uncertainty indication and active learning (AL) algorithms~\cite{lysogorskiy2022active} that can be employed to improve the model if necessary. ACE has successfully been applied to model metallic, covalent and ionic materials and was shown to offer superior accuracy and computational efficiency~\cite{lysogorskiy2021performant, lysogorskiy2022active, qamar2022atomic}.

The ACE methodology combines the advantages of machine-learned (ML) methods and physically based models of interatomic interactions. We summarize only the essentials of ACE and refer to original publications for more details~\cite{drautz2019atomic, drautz2020atomic, dusson2022atomic, lysogorskiy2021performant, bochkarevefficientpara, bochkarev2022multilayer}.

One of the key features of ACE is a complete and hierarchical set of basis functions $\BBb_{i \vii}$ that span the space of local atomic environments. This enables to expand an atomic property $\ace_i^{(p)}$ such as the energy of atom $i$ as
\begin{equation}
\ace_i^{(p)} = \sum_{\vii}^{n_{\vii}} \cBB_{\vii}^{(p)} \BBb_{i \vii} \,, \label{eq:ACEproperty} 
\end{equation}
with expansion coefficients $\cBB_{\vii}^{(p)}$ where ${\vii}$ is composed of several indices.  The basis functions fulfill fundamental translation, rotation, inversion and permutation (TRIP) invariances for the representation of scalar variables, or equivariances for the expansion of vectorial or tensorial quantities. 

In the simplest case, the energy is evaluated linearly as
\begin{equation}
   \label{eq:ACEenergylinear} 
    E_i = \ace_i^{(1)} \, ,
\end{equation}
but most ACE parametrizations to date, including the presented Pt-Rh model, use two atomic properties with a Finnis-Sinclair square-root embedding,
\begin{equation}
E_i =  \ace_i^{(1)} + \sqrt{\ace_i^{(2)}}  \,. \label{eq:EFS}
\end{equation}

We demonstrate that Pt-Rh ACE provides not only an accurate description of fundamental properties for both elemental metals as well as their compounds, but that it is well suited for large-scale atomistic simulations of nanoclusters.

%%%%%%%%%%%%%%%%%%%%%%%%%%%%%%%%%%%%%%%%%%%%%%%%%%%%%%%%%%%%%%%%%%%%%%%%%%%%%%%%%%%%%%%%%%%%%%%%%%%%%%%%%%%%%%%

\section{Results and discussion}\label{sec:results}

\subsection{Training of ACE}\label{sec:training}

\begin{figure}
    \centering
    \includegraphics[scale=0.45]
    {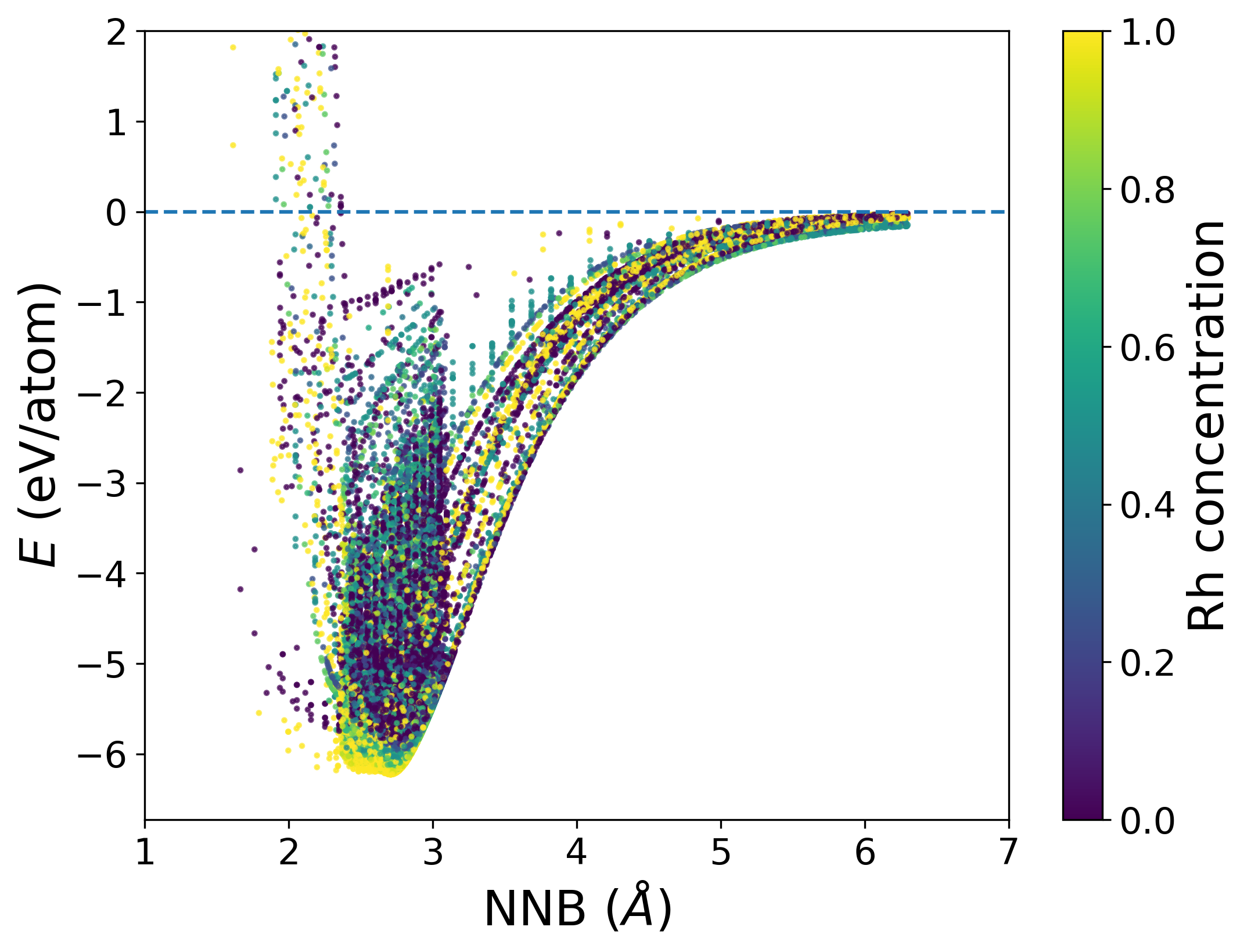}
    \caption{Energy per atom of structures in the Pt-Rh training set as a function of the nearest neighbour distance (NNB).}
    \label{fig:binaryEnn}
\end{figure}

\begin{figure}[htbp]
 \subfigure[]{
    \includegraphics[width=0.48\textwidth]
    {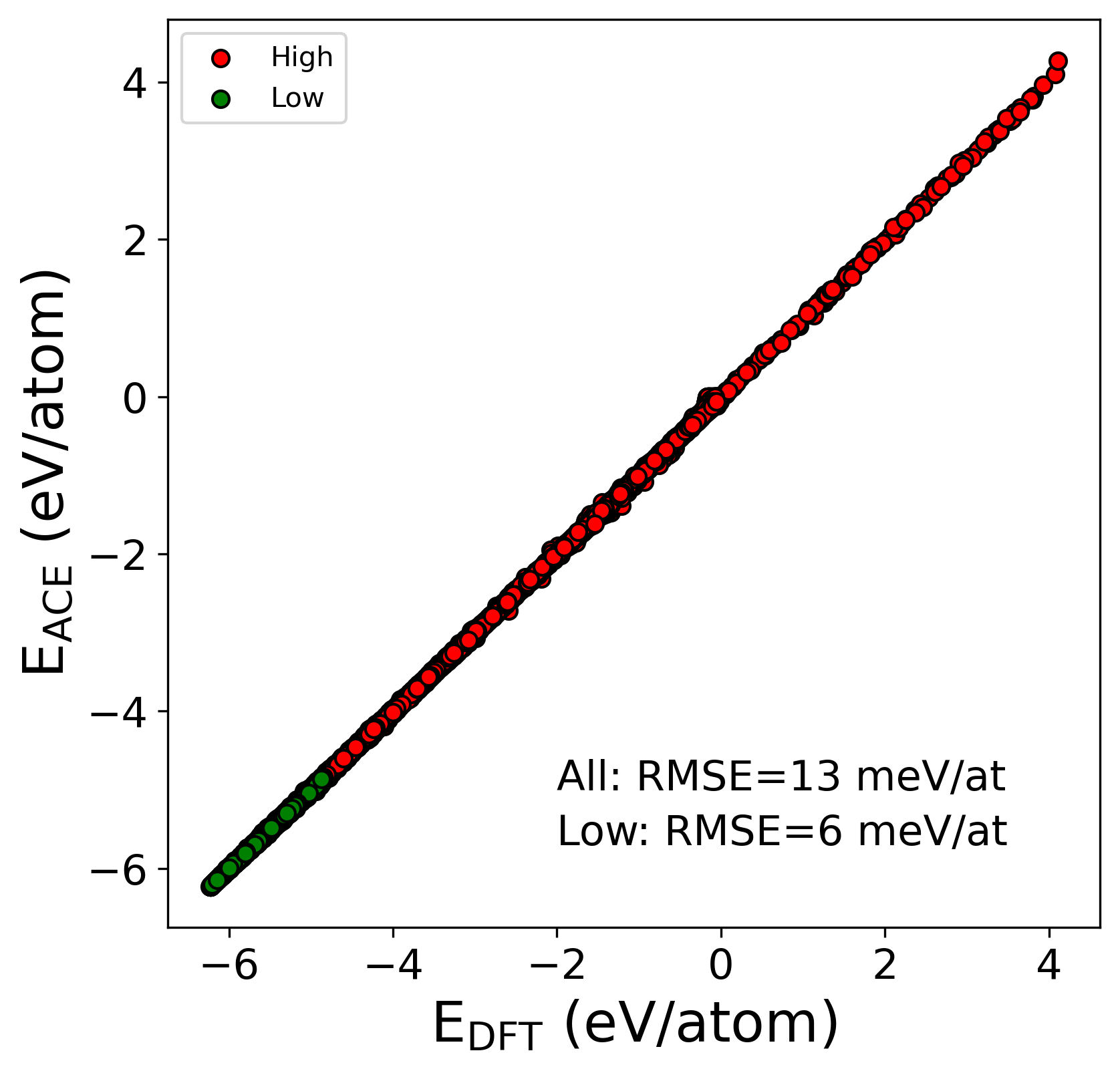}}
 \subfigure[]{
     \includegraphics[width=0.48\textwidth]
     {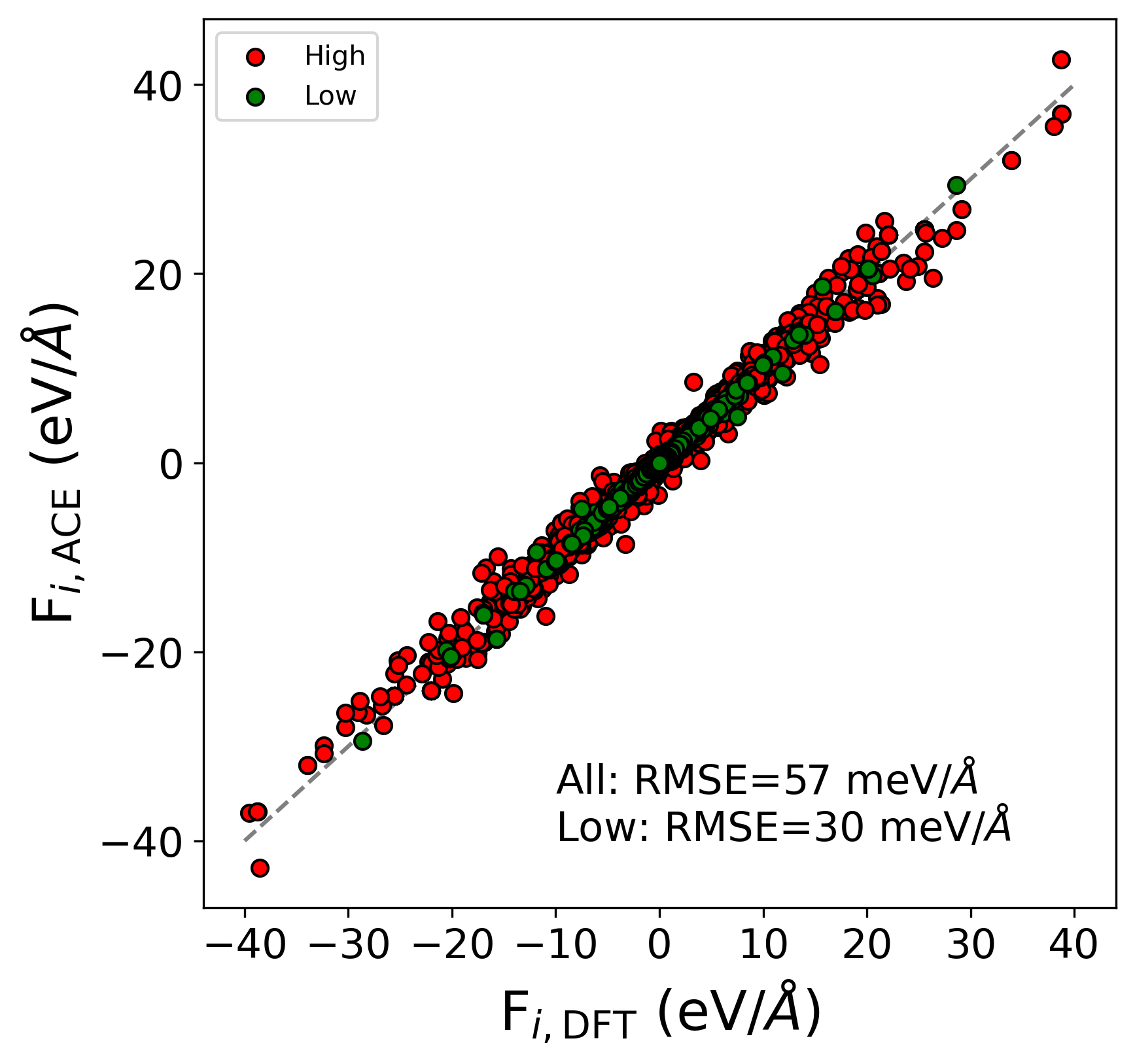}}
     
    \caption{Comparison of energies and forces predicted by ACE with respect to the reference training DFT data.
      The ``Low'' subset includes data within a distance of 1\,eV/atom above the convex hull.
    }
    \label{fig:EFacedft}
\end{figure}

The parameterization of ACE models requires a consistent set of reference DFT data. 
For the Pt-Rh system, we generated about 30 000 atomic configurations and evaluated their energies and forces using the Vienna ab-initio simulation package (VASP)~\cite{kressabinitio1993,kresse1996efficiency,kresse1996efficient} and the PBE functional for approximating exchange and correlation\cite{PBEXC}. The considered structures span a wide portion of the configuration space for both elemental metals and their binary compounds. The configurations include perfect bulk structures, common defects, e.g., vacancies, surfaces and grain boundaries, and small random clusters. For all these configurations, we varied their volumes and randomly displaced atoms away from their equilibrium positions to sample vibrational degrees of freedom. A summary of energies per atom for all structures as a function of the nearest neighbour distance in each structure is shown in Fig.~\ref{fig:binaryEnn}.  The training data corresponds to 29800 structures with 437157 atoms, spanning nearest neighbour distances from 1.8 up to 6.3~\AA\ and sampling the energy space from about $-6.3$~eV/atom to $+2$~eV/atom.

The ACE parameterization was carried out using the \texttt{pacemaker} package~\cite{bochkarevefficientpara}. 
In the fitting process, a hierarchical basis extension scheme with a sequential addition of ACE basis functions was adopted. For the radial basis functions, spherical Bessel functions were used. Structures within an energy window of 1\,eV/atom above the convex hull were assigned significantly larger weights in the loss function optimization, while structures with higher energy were given smaller weights. 

After initial training, we employed AL to ensure a reliable description of Pt-Rh surfaces as these are crucial for simulations of core-shell nanoclusters (see below). 
This was achieved by computing the extrapolation grade $\gamma$ based on the D-optimality criterion~\cite{lysogorskiy2022active} for a number of (100), (110) and (111) fcc surface slabs containing up to 56 atoms with two compositions (Pt$_{0.3}$Rh$_{0.7}$ and Pt$_{0.7}$Rh$_{0.3}$). 
The slab configurations were obtained using a hybrid MD-MC method in the $NVT$ ensemble at three different temperatures (300, 500 and 800\,K).
Structures with large values of the extrapolation grade ($\gamma >5$) were selected using the MaxVol algorithm. These structures were then computed with DFT, added to the training set, and the ACE potential was retrained. See Section~\ref{sec:methods:active_learning} for more details.

In Fig.~\ref{fig:EFacedft}, the energies and forces predicted by ACE are compared with the reference DFT data for the training set. 
The ACE model reproduces the DFT energies and forces with good accuracy. 
Obtained root mean square errors (RMSE) for the energy and force components are 13~meV/atom and 57~meV/\AA\ for the training set and 36~meV/atom and 140~meV/\AA\ for the test set, respectively. The relativley large errors in the energies are due to the very low weights that were assigned to high energy structures with atoms at close distance as these structures have no relevance for the simulation of the nanoparticles. In fact, 
for the lower energy structures within the 1~eV window above the convex hull that are most significant in the simulations, the RMSE are only 6\,meV/atom and 30\,meV/\AA\ for the training set and 6\,meV/atom and 26\,meV/\AA\ for the test set, respectively.

\subsection{Validation of the Pt-Rh ACE}

For an initial validation of the parameterized potential, various fundamental properties of the elemental metals were examined and compared to DFT and experimental data in Table~\ref{tab:diffproperty}. For both Pt and Rh, the equilibrium lattice parameters predicted by ACE are in excellent agreement with DFT, while being slightly larger than the experimental values. The elastic constants and bulk moduli also agree well with those of DFT, with somewhat larger discrepancies observed for $C_{11}$ and $C_{44}$ for Pt. The differences however remain comparable to those between DFT and experiment. The larger lattice constants and mostly smaller elastic moduli predicted by DFT and ACE in comparison with experiment are likely caused by the PBE-GGA exchange-correlation functional that leads to underbinding for nonmagnetic fcc metals~\cite{graboskiabinitio}.

The formation and migration energies of the mono-vacancy are in close agreement with the reference DFT values and comparable with the available experimental measurements for Rh. For Pt, the large discrepancy between the theoretical and experimental values is due to deficiencies in the description of the exchange-correlation contributions at metallic surfaces~\cite{thomasvacancy}. A better agreement can be obtained after correcting for this intrinsic surface error, but the correction term depends on the electronic structure of the metal~\cite{thomasvacancy} and cannot be employed in classical atomistic models.

\begin{figure}
    \centering
    \includegraphics[scale=0.45]{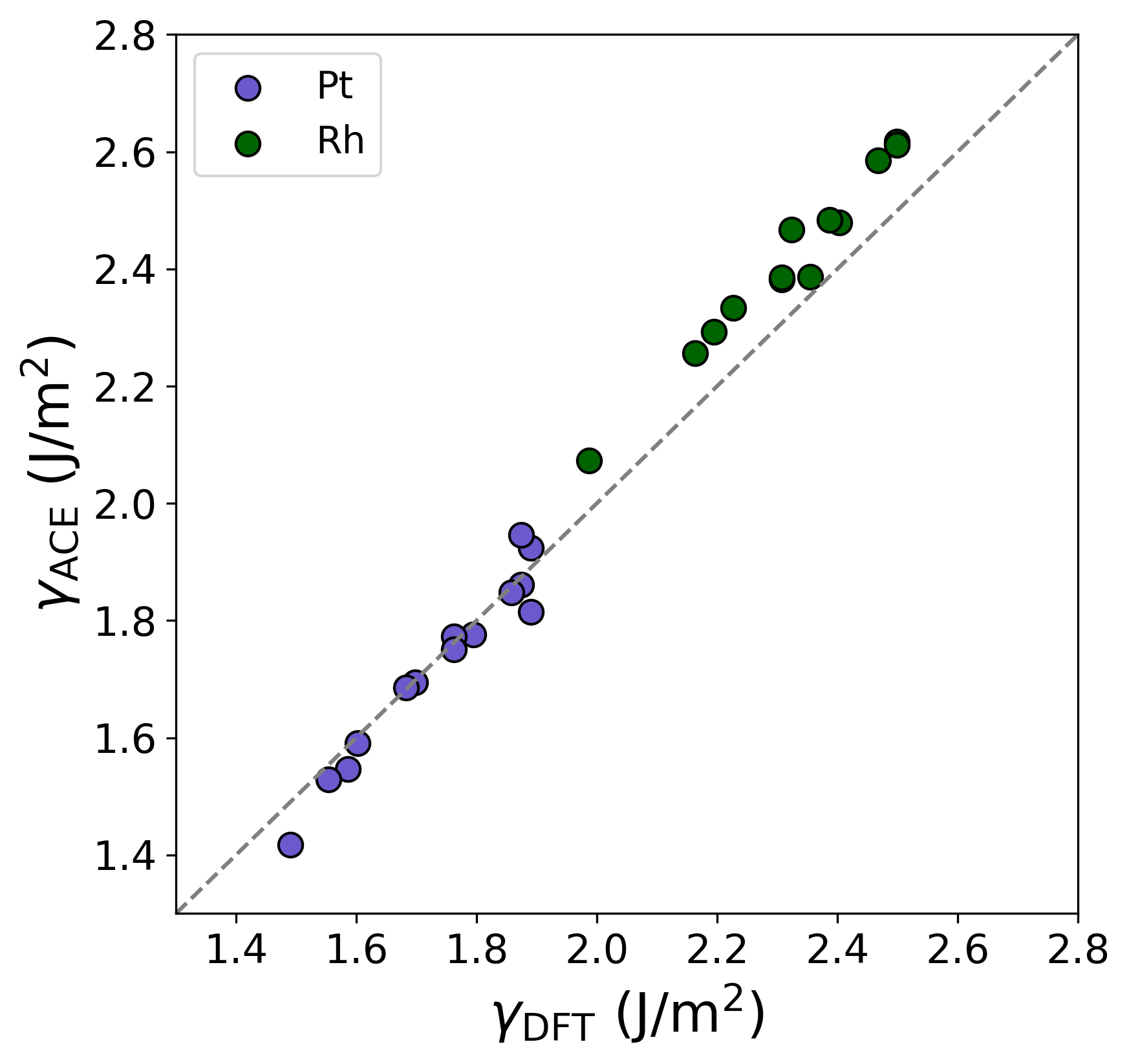}
    \caption{Comparison of various surface energies obtained from ACE and DFT. The reference DFT data are taken from the Crystalium database\cite{tran2016surface,tran2019anisotropic,zheng2020grain}. %\textcolor{orange}{values updated for upfitted most recent pot.}
    }
    \label{fig:surfEcomparison}
\end{figure}

We also examined energies of surfaces with low-index orientations. As illustrated in Table~\ref{tab:diffproperty}, ACE captures correctly the ordering of Pt surface energies in comparison to DFT, as $\gamma_{(100)}>\gamma_{(110)}>\gamma_{(111)}$. However,  DFT strongly underestimate the absolute values of surface energies in comparison to experiment. The underlying reason for this discrepancy is identical as in the case of vacancies. For Rh, the stability of the lowest energy $(111)$ surface is reproduced by ACE, but the energies of the $(100)$ and $(110)$ are almost degenerate according to DFT while ACE predicts the latter to have a slightly higher energy. Besides these common surface orientations, we also evaluated other surface configurations provided in the Crystalium database\cite{tran2016surface,tran2019anisotropic,zheng2020grain} and compared them to the available DFT data in Fig.~\ref{fig:surfEcomparison}. A close agreement is seen between ACE and DFT with a slight overestimation of Rh surface energies by ACE. When comparing both metals, it is evident that Pt has smaller surface energies than Rh, which is relevant for the structure of nanoclusters (see below).

\begin{table}[h]
\centering
\caption{Comparison of various properties of elemental Pt and Rh obtained from ACE, DFT and experiment. For the vacancy, $E_{f}$ and $E_{m}$ represent the formation and migration energies, respectively. For surface energies, the experimental values are for average orientations.%\textcolor{red}{the exp values of surf should be checked for catalyst.}
}
\begin{tabular}{lccc|ccc}
\toprule
&\multicolumn{3}{c}{Platinum}&\multicolumn{3}{c}{Rhodium}\\
%& & Ag &  & &Pd &\\
 & ACE & DFT & Exp & ACE & DFT & Exp\\
\midrule
Lattice constant & & & &&&\\
$a_0$~(\AA) & 3.97 & 3.97 & 3.92\cite{waseda1975high} &3.83 & 3.83 & 3.80\cite{nuding1997influence}\\ 
\midrule
%\midrule 
Elastic moduli & &&&&&\\
$C_{11}$~(GPa) & 267  & 308 & 347\cite{simmons1965single}  &422  & 409   & 422\cite{simmons1965single} \\
$C_{12}$~(GPa) &227  & 222 &251\cite{simmons1965single}& 169 & 182  & 192\cite{simmons1965single}\\
$C_{44}$~(GPa) &50  & 70 & 77\cite{simmons1965single} &192 & 186 &  194\cite{simmons1965single}\\
$B$~(GPa) & 241 & 251 & 228$-$275\cite{darling1966anon} & 253& 258 & 276\cite{buch1999pure}\\
\midrule
Vacancy &&&&&&\\
$E_{f}$~(eV) &0.69 &0.68\cite{thomasvacancy} &1.15$-$1.60\cite{cahn1996physical} &1.84 &1.82 &1.71\cite{de1988cohesion}\\
$E_{m}$~(eV) &1.06 &1.24\cite{angsten2014elemental} &1.43\cite{balluffi1978vacancy} &1.87 &1.79\cite{angsten2014elemental} &--\\
\midrule
Surface energy &&&&&&\\
$\gamma_{(100)}$~(mJ/m$^2$) & 1847 & 1856\cite{Ptosti_1189002} &  & 2386& 2350\cite{Rhosti_1287920} & \\
$\gamma_{(110)}$~(mJ/m$^2$) & 1685 & 1681\cite{Ptosti_1189002} & 2490\cite{tyson1977surface} & 2466 & 2331\cite{Rhosti_1287920} & 2659\cite{tyson1977surface}\\
$\gamma_{(111)}$~(mJ/m$^2$) & 1417 & 1488\cite{Ptosti_1189002} &  & 2073 & 1984\cite{Rhosti_1287920} & \\
\bottomrule
\end{tabular}
\label{tab:diffproperty}
\end{table}

\begin{figure}[htpb]
\centering
   \subfigure[Pt]{
   \includegraphics[width=0.48\textwidth]{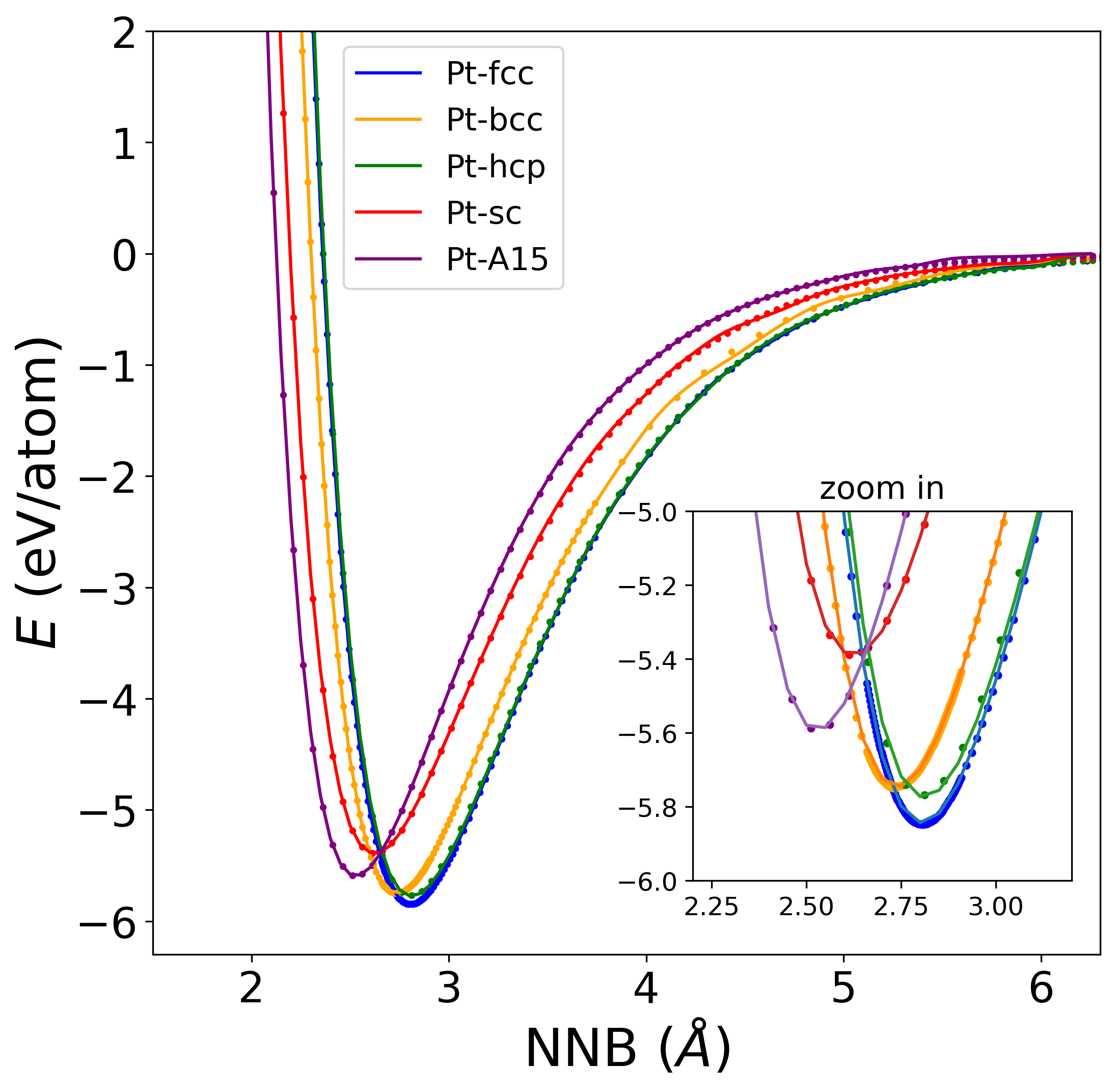}
   }
   \subfigure[Rh]{
   \includegraphics[width=0.48\textwidth]{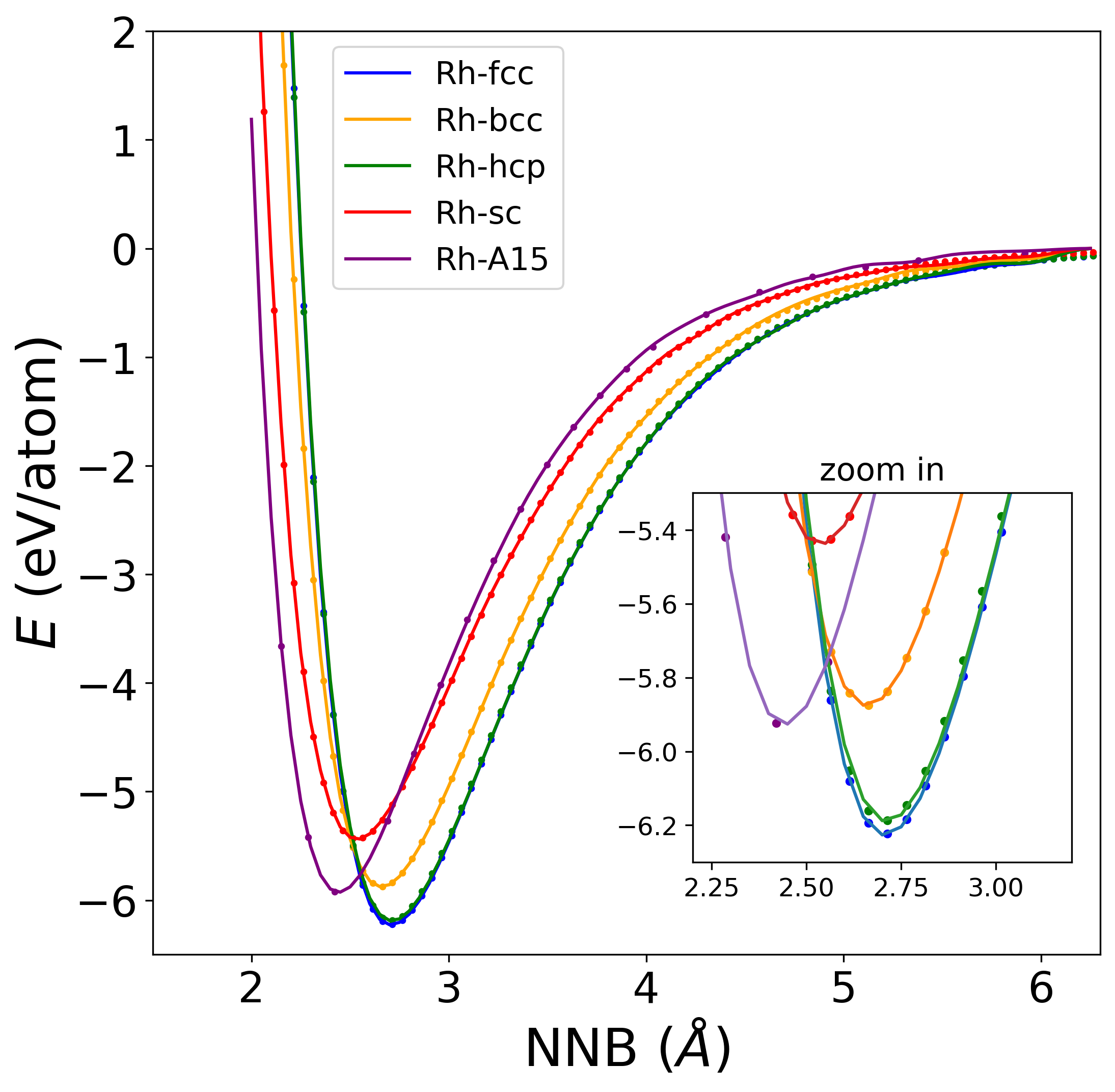}}
    \caption{Energy as a function of the nearest neighbour distance (NNB) for various bulk phases of (a) Pt and (b) Rh. The lines represent ACE predictions while the dots correspond to the reference DFT data. The insets depict low-energy regions around the fcc ground state.}
    \label{fig:Ennacedft}
\end{figure}

The phase stabilities of various bulk phases over a broad range of volumes are examined for both metals in Fig.~\ref{fig:Ennacedft}. It can be seen that the agreement between ACE and DFT is excellent up to the dissociation limit, which is often not the case for ML models~\cite{lysogorskiy2022active, qamar2022atomic}. Both Pt and Rh are most stable in the fcc phase. The closely related hexagonal close-packed (hcp) phases have only slightly higher energies ($\Delta E_\text{hcp-fcc}=79$~meV/atom for Pt and $\Delta E_\text{hcp-fcc}=36$~meV/atom for Rh). The ACE model correctly reproduces the energetic ordering of other low-energy phases, namely, $E_\text{sc}>E_\text{A15}>E_\text{bcc}>E_\text{hcp}>E_\text{fcc}$ for Pt and $E_\text{sc}>E_\text{bcc}>E_\text{A15}>E_\text{hcp}>E_\text{fcc}$ for Rh, where the subtle differences of $E_{\text{bcc}-\text{hcp}}=18$~meV/atom for Pt and $E_{\text{bcc}-\text{A15}}=49$~meV/atom for Rh are all correctly captured and bcc and sc correspond to body-centered cubic and simple cubic phases, respectively.

\begin{figure}[htpb]
\centering
   \subfigure[]{
    \includegraphics[width=0.48\textwidth]{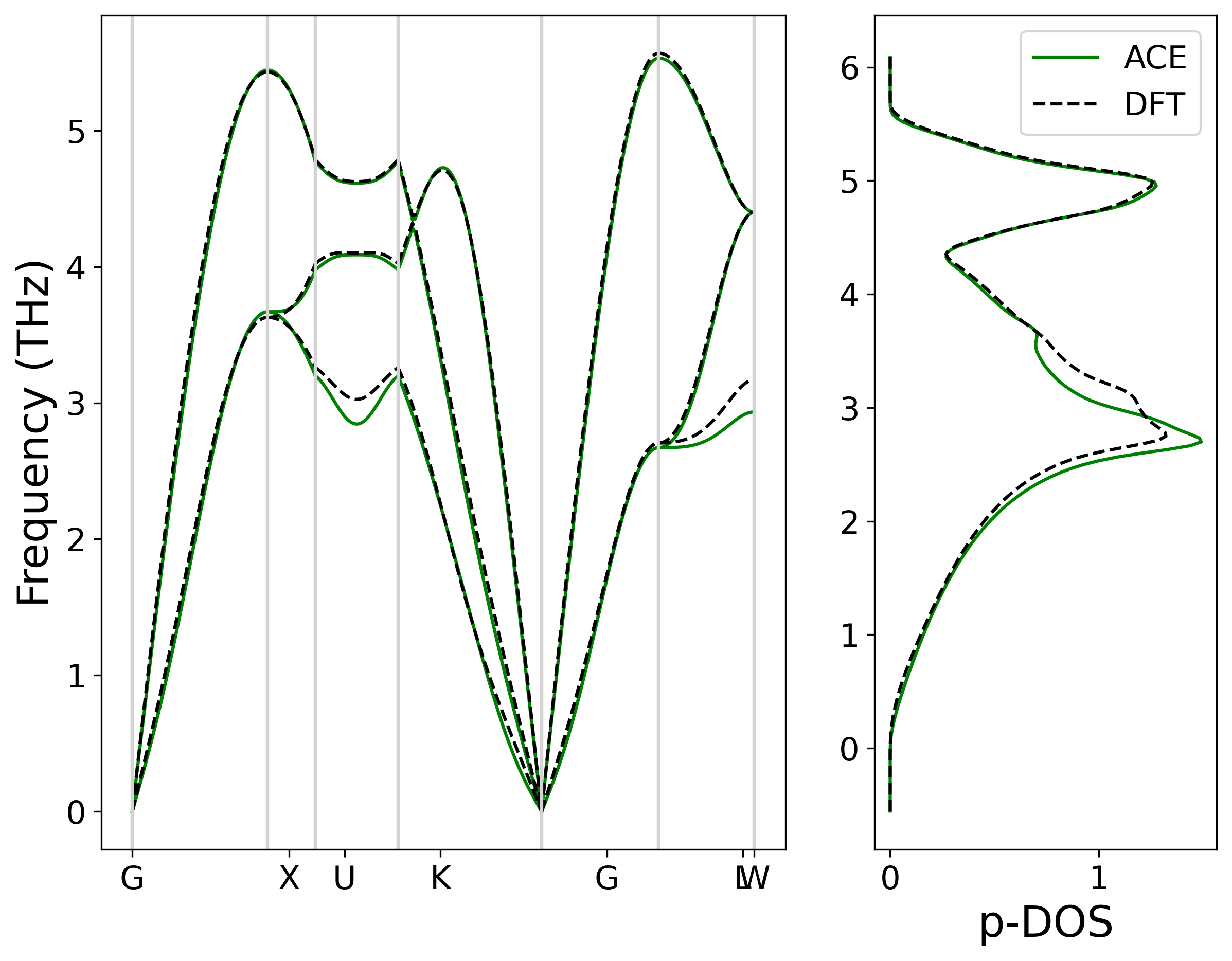}}
   \subfigure[]{
   \includegraphics[width=0.48\textwidth]{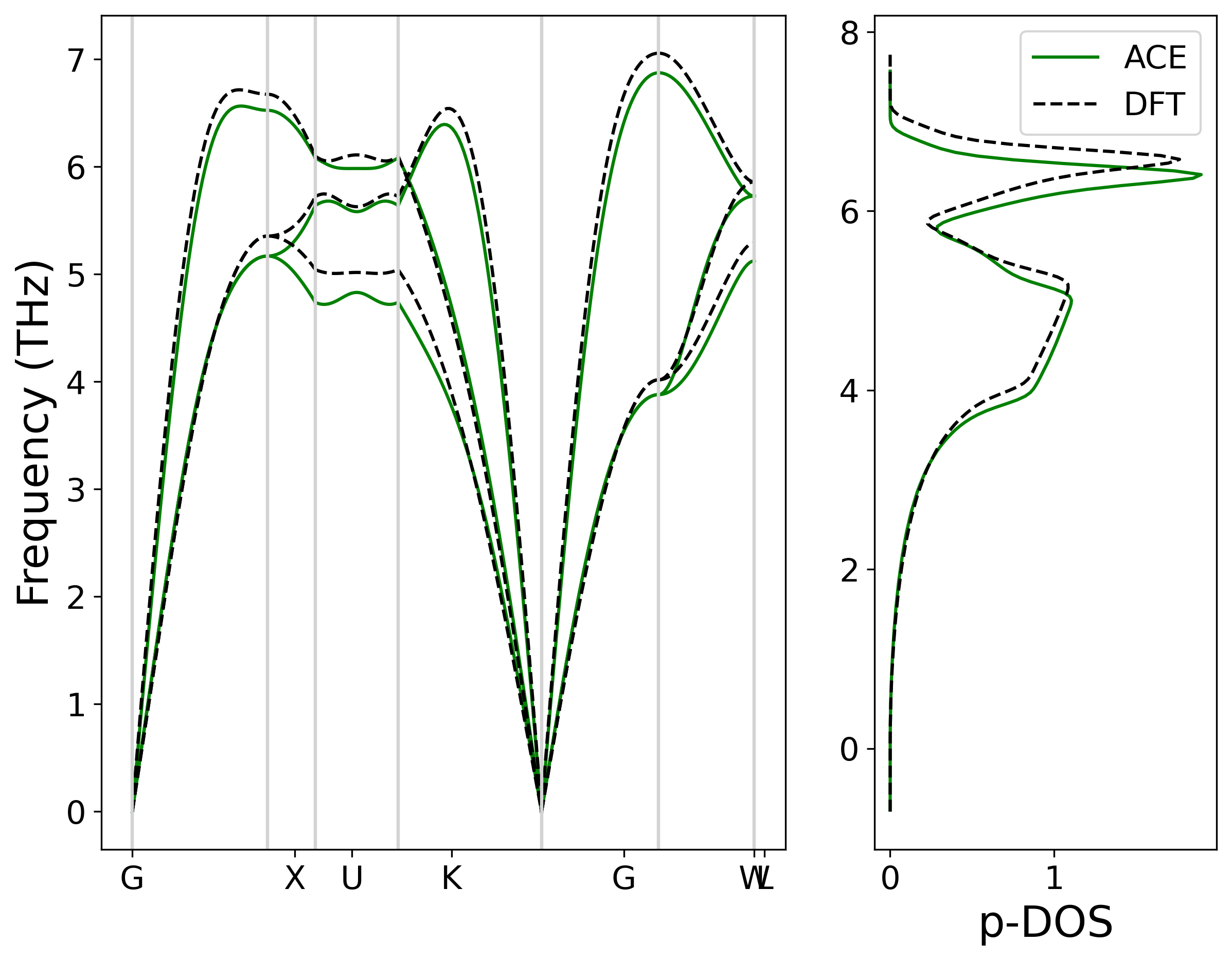}}
    \caption{Phonon dispersion for the ground state fcc structure of (a) Pt and (b) Rh.}
    \label{fig:phonons}
\end{figure}

Phonon dispersions along high symmetry directions of the Brillouin zone and phonon densities of states for the fcc phases of Pt and Rh are shown in Fig.~\ref{fig:phonons}. ACE exhibits a close agreement with DFT for both elemental metals.

\begin{figure}[htbp]
    \centering
    \subfigure[]{
    \includegraphics[width=0.45\textwidth]{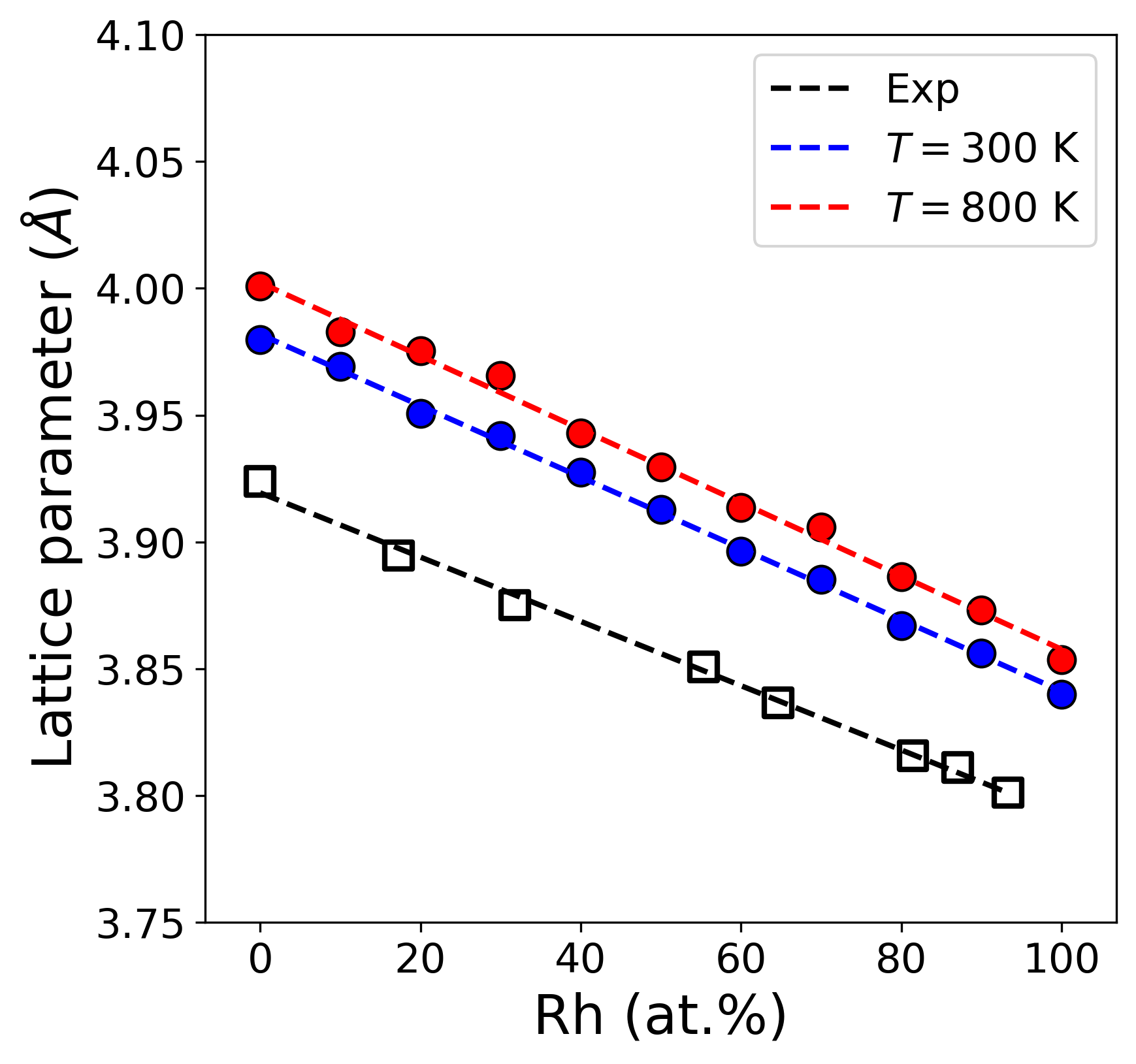}}
    \subfigure[]{
    \includegraphics[width=0.45\textwidth]{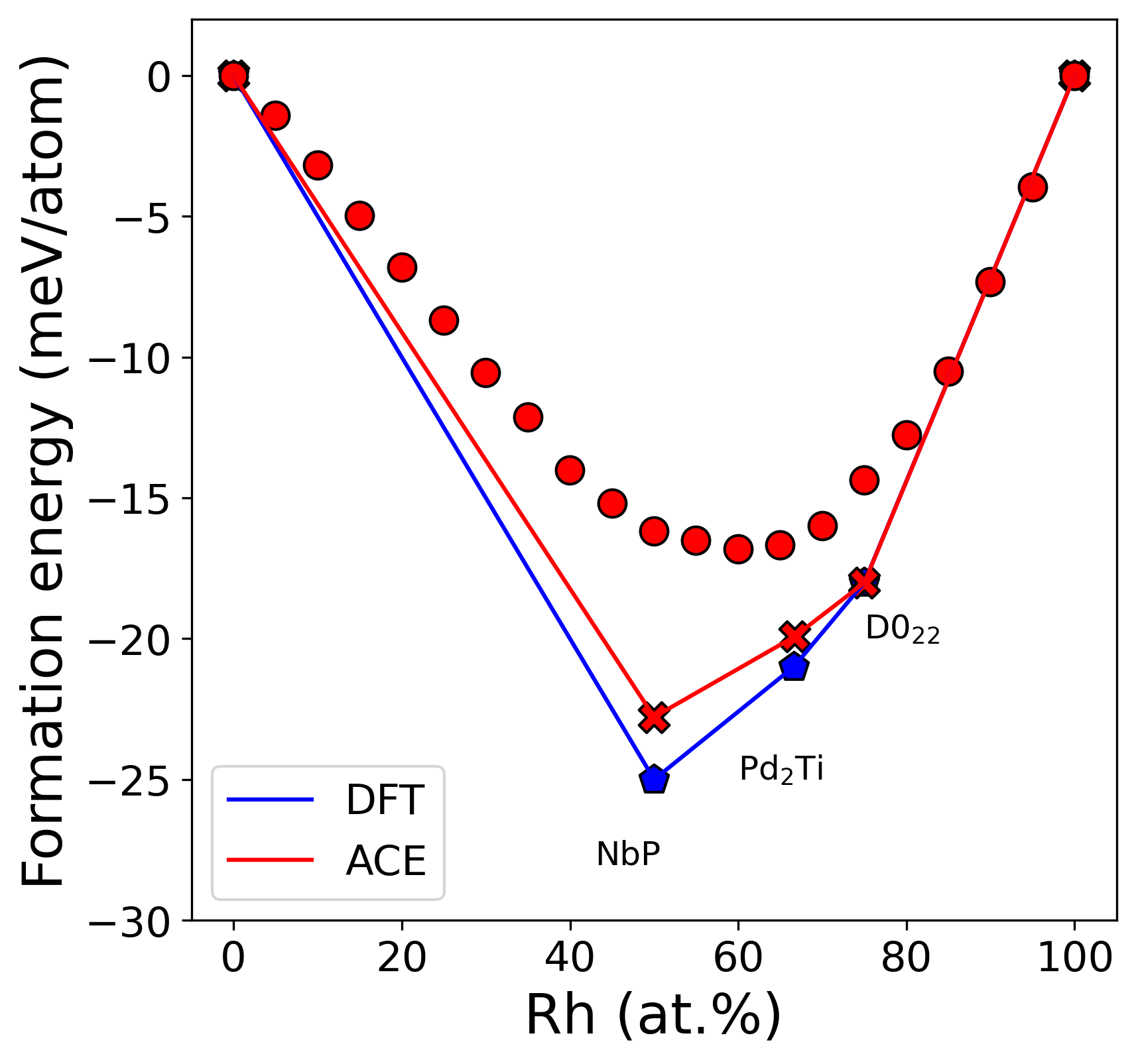}}
    \caption{(a) Lattice constants of Pt-Rh alloys as a function of chemical composition at different temperatures. The square symbols represent the experimental values at $T=300$~K \cite{raub1959metals}. (b) Formation energies of most stable ordered phases from DFT (blue diamonds)~\cite{hartcomprehensive} and their ACE predictions (red crosses). ACE predictions for fcc random solid solutions are marked by red circles.
    }
\label{fig:latticediffTandEform}
\end{figure}

Apart from the properties of the elemental metals, we evaluated the equilibrium lattice parameters of random Pt-Rh fcc alloys as a function of temperature and compared them with available experimental data in Fig.~\ref{fig:latticediffTandEform}(a). The linear decrease of the lattice parameter with increasing Rh concentration is well reproduced by ACE. As mentioned above, the overestimation of the values of the lattice parameters predicted by theory are due to the underbinding of the PBE exchange-correlation functional used for the DFT training data.

There exists experimental evidence for ordering tendencies in the Pt-Rh system \cite{steiner_PhysRevB.71.104204} but no unequivocal confirmation of the existence of distinct ordered phases. Nevertheless, several theoretical studies~\cite{turchi_PhysRevB.74.064202,curtarolo2005accuracy,lu1991long,lu1995ordering,pohl2009phase} predicted a possible occurrence of ordered phases at low temperatures. The D0$_{22}$ phase and the so-called "phase 40" (NbP prototype, Pearson symbol tI8, space group I4$_1$/amd) were identified as the lowest energy phases for the PtRh$_3$ and PtRh stoichiometries, respectively. More recently, Hart et al. \cite{hartcomprehensive} reported the Pd$_2$Ti prototype (Pearson symbol oI6, space group Immm) to be a stable phase on the convex hull for the PtRh$_2$ stoichiometry. 

We calculated the formation energies of these ordered structures using ACE and compared them with the reference DFT data in Fig.~\ref{fig:latticediffTandEform}~(b). In addition, the plot contains formation energies of fcc random solid solutions.  ACE predictions for the three ordered phases agree closely with the DFT results, reproducing correctly the convex hull. 
All random solid solutions lie above the convex hull, but their formation energies remain negative. 

\subsection{Simulations of Pt-Rh nanoclusters}\label{sec5}

As the Pt-Rh nanoclusters appear to be terminated predominantly with (111) and (100) planes (see Fig.~\ref{fig:Expnano}), we started our simulations using truncated octahedrons that contain six (100) and eight (111) planes. The initial focus of the simulations was to examine the surface segregation that may play a role in stabilization of the core-shell geometry. We investigated 
two compositions of $\text{Pt}_{30}\text{Rh}_{70}$ and $\text{Pt}_{45}\text{Rh}_{55}$, with a random initial distribution of both species in the nanoclusters (see supplementary Fig.~S3).
The simulations were done at $T=500$\,K using a hybrid MD/MC procedure for efficiently reaching thermodynamic equilibrium. 

\begin{figure}[htpb]
    \centering
    \subfigure[$\text{Pt}_{30}\text{Rh}_{70}$]{\includegraphics[width=0.32\textwidth]{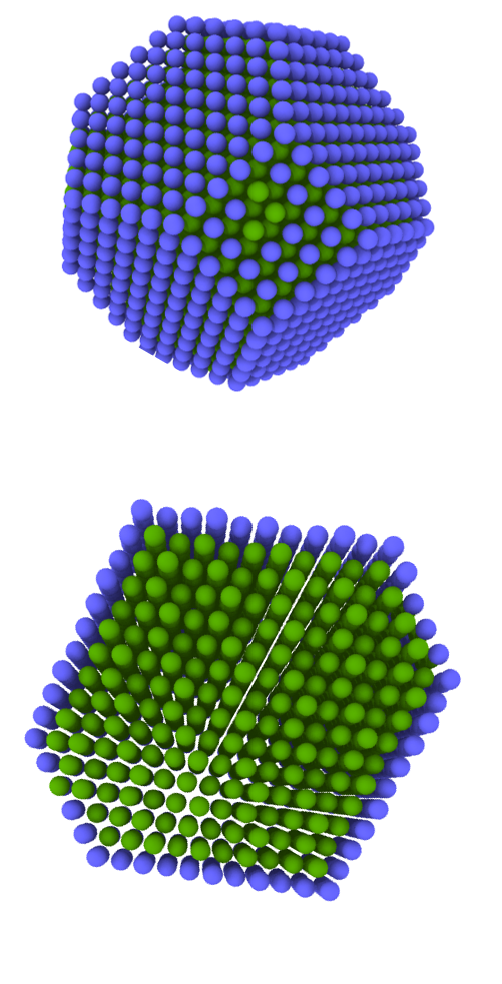}}
    \hspace{1.8 cm}
    \subfigure[$\text{Pt}_{45}\text{Rh}_{55}$]{\includegraphics[width=0.345\textwidth]{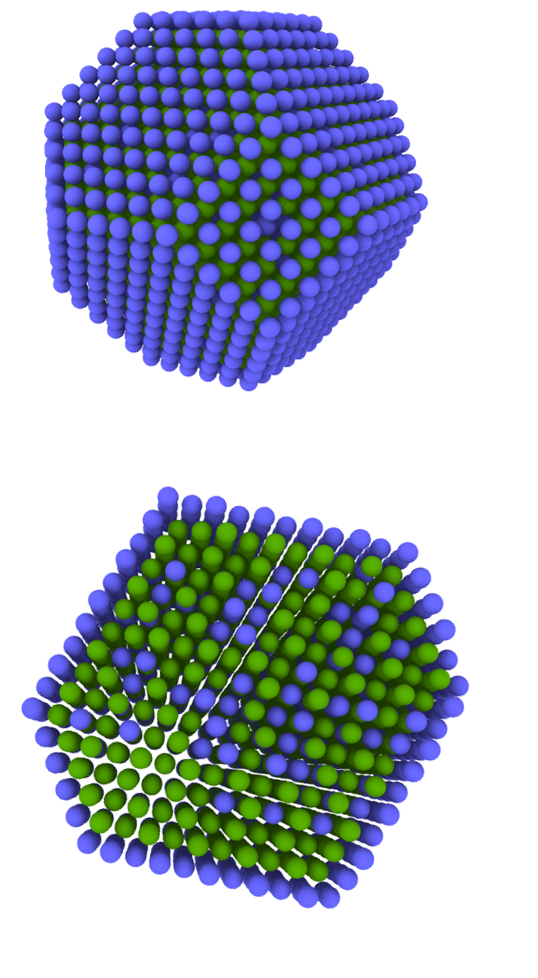}}\\
    \caption{Visualization of the surface segregation in Pt-Rh octahedron nanoclusters with (a) $\text{Pt}_{30}\text{Rh}_{70}$  and (b) $\text{Pt}_{45}\text{Rh}_{55}$ composition. Pt atoms are colored in blue and Rh atoms in green.  
    Top figures show the front view of truncated octahedron with four (111) and one (100) surface facets. Bottom figures show mid cross-sections of the clusters with four (111) and two (100) surface planes. }
    \label{fig:PtRhsurfseg}
\end{figure}

Figure~\ref{fig:PtRhsurfseg} shows the final cluster configurations for both compositions. In both cases, we observed a strong tendency of the Pt atoms (blue spheres) to segregate to the surface and to form a continuous monolayer across the whole cluster surface. Based on the results for the Pt-poor composition $\text{Pt}_{30}\text{Rh}_{70}$, we surmise that the segregation of Pt occurs preferentially on the (111) facets as the coverage of the (100) facets remains incomplete (see the top left configuration in Fig.~\ref{fig:PtRhsurfseg}). For the Pt-rich composition $\text{Pt}_{45}\text{Rh}_{55}$, all surfaces of the cluster are covered completely by Pt (see the top right configuration in Fig.~\ref{fig:PtRhsurfseg}).
Interestingly, the excess Pt atoms in this case do not continue to segregate to adjacent subsurface layers but remain randomly dispersed in the cluster interior. While the formation of a Pt surface layer can be related to the low surface energies of Pt (cf.  Fig.~\ref{fig:surfEcomparison}), experimental observations of the core-shell particles show a thicker coverage of 3-5 layers of Pt at the surface~\cite{vega2023}. However, according to our simulations Rh atoms have a slight preference for the occupation of the second subsurface layer (see the bottom right configuration in Fig.~\ref{fig:PtRhsurfseg}). 

To analyze this discrepancy in more detail, we performed additional DFT calculations of Pt-Rh surface slabs with different orientations. All slabs consisted of ten atomic planes with two Pt and eight Rh layers. One of the Pt layers was located at the surface while the position of the second Pt layer was varied, from being the adjacent subsurface layer, i.e., forming two Pt layers on the surface to the opposite surface of the slab, i.e., a Rh slab with both surfaces covered by Pt layers. These calculations  confirm that the single Pt layers on both slab surfaces corresponds to energetically most favorable configuration, see supplementary Fig.~S4. In contrast, the formation of two adjacent Pt surface layers results mostly in configurations with higher energies than when the Pt layers are farther apart. The energetics of these interactions is described well by ACE, confirming the reliability of ACE predictions in the large-scale MD simulations.  This analysis shows that in thermodynamic equilibrium a multilayer segregation of Pt atoms on the surface of the Pt-Rh nanoclusters is likely not favorable. This suggests that the core-shell morphologies with a Rh core and a thicker Pt shell observed in some STEM experiments are kinetically stabilized. This corroborates with the fact that the experimental synthesis procedure creates the core first followed by the coating of the surface element.

\iffalse
\begin{figure}[htpb]
    \centering
    \subfigure[]{\includegraphics[width=0.35\textwidth]{image/Pt30Rh70_final_anotherview.png}}
    \subfigure[]{\includegraphics[width=0.35\textwidth]{image/Pt30Rh70_final_cut.png}}
    \caption{Visualization of the surface segregation in $\text{Pt}_{30}\text{Rh}_{70}$. Pt atoms are colored in blue and Rh atoms are in green. (a) Surface segregation of Pt in (100) and (111) planes. (b) A cut of the nanocluster illustrating the occupation of Pt at the surface layer. \yan{Pt45 also here. top full, bottom the semi-transparent Rh(whole cluster without the deletion of surface layer).}}
    \label{fig:Pt30Rhsurfseg}
\end{figure}
\fi

To obtain a complementary view on the stability of the core-shell clusters, we performed additional MD simulations at elevated temperatures. In contrast to the hybrid MD-MC simulations presented above, we generated the initial clusters as spheres with a Rh core and a Pt shell, to mimic real experimental geometries. The thickness of the shell was set to approximately three Pt layers. The clusters were then annealed at temperatures of 1000 and 1500 K.

\begin{figure}[htpb]
    \centering
    \subfigure[$T=1000$\,K, $t=2$\,ns]{\includegraphics[width=0.45\textwidth]{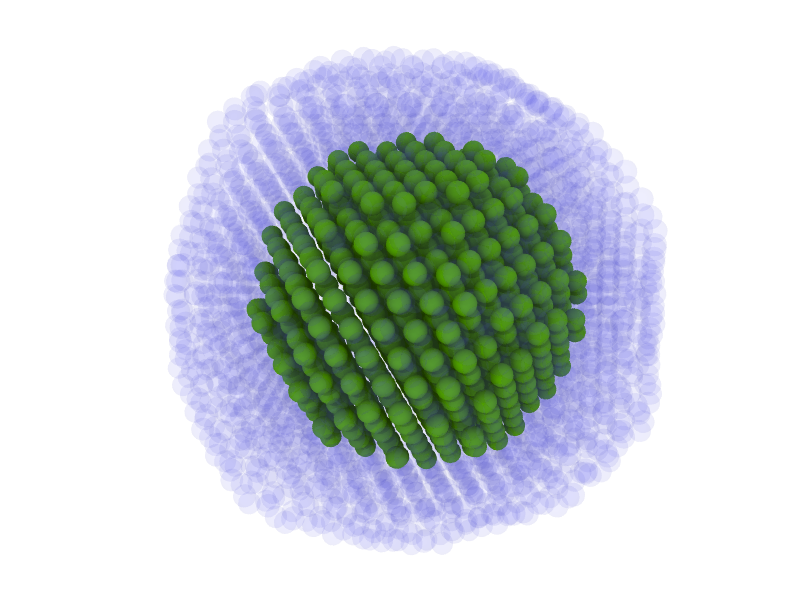}}
    \subfigure[$T=1500$\,K, $t=2$\,ns]{\includegraphics[width=0.45\textwidth]{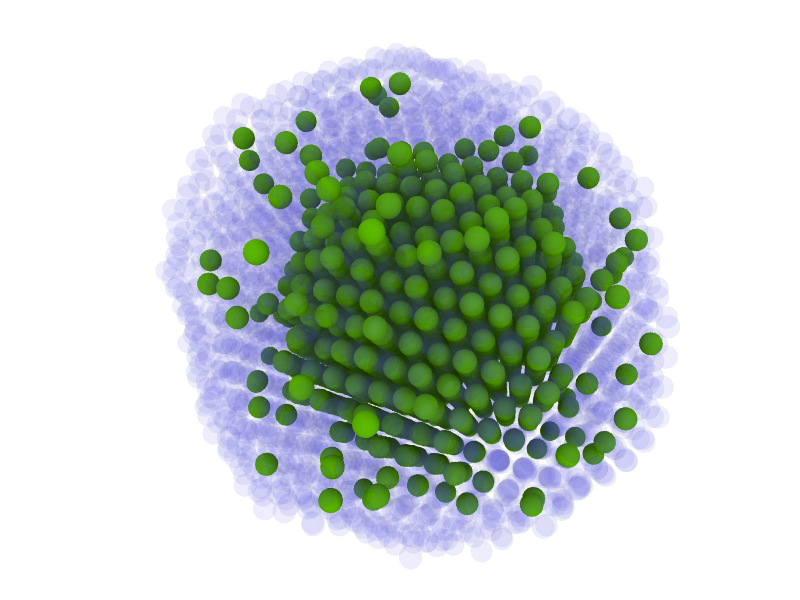}}\\
    \caption{Visualization of core-shell nanocluster after $t=2$~ns annealing at (a) 1000 K and (b) 1500 K. Rh atoms in green, Pt atoms in transparent blue.}
    \label{fig:PtRhaneal}
\end{figure}

In Fig.~\ref{fig:PtRhaneal}, the nanoclusters are visualized after 2 ns of annealing. At the lower temperature of 1000 K, the cluster undergoes small shape changes but its core-shell morphology remains intact. When the temperature is increased to 1500 K, we start observing diffusion events leading to a gradual intermixing of the Rh atoms from the core into the Pt shell. The duration of the MD simulation is too short to reach thermodynamic equilibrium and supports our conclusion that the experimentally observed Rh core and Pt shell structure is metastable.

%%%%%%%%%%%%%%%%%%%%%%%%%%%%%%%%%%%%%%%%%%%%%%%%%%%%%%%%%%%%%%%%%%%%%%%%%%%%%%%%%%%%%%%%%%%%%%%%%%%%%%%%%%%%%%%

\section{Conclusion}\label{sec6}

We presented the development and application of ACE for the Pt-Rh binary system. Our approach utilized a semi-automatic workflow combining high-throughput DFT calculations, efficient fitting and validation procedures, and active learning using uncertainty indication based on the D-optimality criterion.  The resulting ACE potential accurately describes fundamental properties of the Pt-Rh system with ab initio accuracy and is suitable for large-scale atomistic simulations.

We applied the ACE to investigate structural stability of various Pt-Rh nanoclusters using hybrid MC/MD simulations. Our results show a strong surface segregation of Pt and a preference to form a single monolayer coverage of the whole cluster. This is in contrast to experimental observations, where core-shell nanoclusters with thicker Pt shells were reported. By carrying out MD simulations at elevated temperatures, we showed that the core-shell cluster morphologies consisting of a Rh core with a thicker Pt shell observed in experiments are not thermodynamically favorable but rather kinetically stabilized. 

The accuracy of ACE cannot exceed the accuracy of the reference DFT calculations. DFT predicts surface and defect energies in Pt poorly, which suggests that the simulations for Pt-Rh nanoclusters are limited primarily by the DFT reference data. This highlights that improved exchange correlation functionals that are suitable for high-ghroughput database generation are urgently required.

%%%%%%%%%%%%%%%%%%%%%%%%%%%%%%%%%%%%%%%%%%%%%%%%%%%%%%%%%%%%%%%%%%%%%%%%%%%%%%%%%%%%%%%%%%%%%%%%%%%%%%%%%%%%%%%

\section{Methods}

\subsection{DFT calculations}

All reference DFT calculations were done using the Vienna ab-initio simulation package (VASP)~\cite{kressabinitio1993,kresse1996efficiency,kresse1996efficient} with projector augmented waves~\cite{Blochlprojectoraugemented}. For the description of the exchange-correlation functional, we employed the Perdew-Burke-Ernzerhof (PBE) functional \cite{PBEXC}. For energy and force calculations parameters were chosen with a tight settings to obtain an converged accurate results. A plane-wave cutoff energy of $E_\text{cut}=500$~eV and Gaussian smearing of $\sigma=0.1$~eV were used in all calculations. A dense $\Gamma$ centered k-point mesh with a spacing of 0.125 \AA$^{-1}$ was used to sample the Brillouin zone. 

\subsection{ACE parametrization}

The ACE model was parameterized using the \texttt{pacemaker} package~\cite{bochkarevefficientpara}. 
In the first stage, fitting was performed using a hierarchical basis extension, resulting in the ACE potential consisting of 8874 basis functions and 20 688 parameters with a maximum order $\nu=4$, corresponding to five-body interactions. Additional details of the model are summarized in Table~\ref{tab:ace_shape}. Relative weights of different structures in the loss function were distributed based on their distance to the convex hull. 
This method prioritized structures with lower energies, giving them a greater contribution to the loss function~\cite{bochkarevefficientpara}. In addition, the weights of non-periodic clusters were reduced by a factor of ten to provide regularization to the ACE model.
The initial parametrization was improved using active learning as described in Section~\ref{sec:methods:active_learning}.

\begin{table*}[h!]
\caption{Details of the ACE parametrization: $r_{c}$ denotes the cutoff radius; RBF the type of radial basis function; "\#func. /element" represents the maximum number of functions per element for each body order $\nu$.}
\label{tab:ace_shape}
\begin{tabular}{cccccc}
\toprule
System & $r_{c}$ (\AA)  & RBF   & $n_\text{max}$     & $l_\text{max}$    & \#func. /element   \\
\midrule
Pt-Rh  & 6.3 & SBessel  & 28/5/4/1 & 0/6/4/1 & 56/385/3976/20   \\
\bottomrule
\end{tabular}
\end{table*}

\subsection{MD and MC simulations}

The Large-scale Atomic/Molecular Massively Parallel Simulator (LAMMPS) package~\cite{LAMMPS} with the ML\_PACE~\cite{lysogorskiy2021performant} and MC~\cite{sadighscalable} packages was used for the MD and MC atomistic simulations.  We adopted an integration time step of $\Delta t=1$~fs for the MD simulations. All simulations were equilibrated at the target temperature/pressure for $t_{eq}=50$~ps before production runs. For MD simulations of lattice parameters %and densities 
of Pt-Rh alloys, we used a simulation cell containing 2048 atoms that corresponds to an $8\times8\times8$ conventional fcc cell. 
Simulations were conducted at two temperatures $T=300$\,K and $T=800$\,K in an isothermal-isobaric ($NPT$) ensemble with a pressure of $P=0$~bar for $t=500$~ps. 
Nos\'{e}-Hoover thermostat and barostat were used to control temperature and pressure. 
Damping parameters were  0.1\,ps for temperature and 1\,ps for pressure to ensure the temperature and pressure fluctuations are appropriate. 

Two nanocluster geometries were created using the Atomistic Simulation Environment (ASE)~\cite{ase-paper}:
a truncated octahedron for the surface segregation phenomena and a spherical nanocluster for annealing of nanoclusters. The truncated octahedron comprises six $(100)$ and eight $(111)$ surfaces and contained 2406 atoms. Different chemical compositions were initialized by randomly assigning Pt or Rh to the atomic positions. A spherical nanocluster was created with 3577 atoms consisting of a Rh core (1058 atoms) covered by three layers of Pt  (2519 atoms).  All simulations of nanoparticles were performed in the canonical ($NVT$) ensemble using the Nos\'{e}-Hoover thermostat with a damping parameter of 0.1\,ps.

A Monte Carlo scheme was used for the simulation of surface segregation using the MC~\cite{sadighscalable} packages implemented in LAMMPS to allow for atomic swapping between different types of atoms. The simulations were performed for $t=200$\,ps for surface segregation and $t=2$\,ns for annealing of nanoparticles. The open visulization tool OVITO~\cite{stukowski2009visualization} was used for the visulization of nanoclusters.

\subsection{Active learning}
\label{sec:methods:active_learning}

To capture atomic environments relevant for nanoclusters, we used small fcc surface structures with different crystallographic orientations and sizes, namely, slabs with the (100) surface orientation containing 32 atoms, the (110) surface of 56 atoms, and (111) surface orientation containing 36 atoms, respectively.
To sample different slab configurations, we employed a hybrid MD-MC approach in the $NVT$ ensemble, running each simulation for 100 ps at three different temperatures (300, 500 and 800 K). 
We calculated on-the-fly the extrapolation grade $\gamma$ for each atom every 100th simulation step using the D-optimality criterion. Structures with $\gamma$ values greater than 5 were selected as extrapolative structures, resulting in a few thousand candidate structures. All simulations were performed using the LAMMPS version, which calculates the extrapolation grades for ACE~\cite{lysogorskiy2022active}.
To select the most representative atomic environments, we utilized the MaxVol algorithm to construct a new active set from the structures generated in the previous step. We gave priority to structures that had multiple atoms that entered the new active set~\cite{lysogorskiy2022active} and selected the top hundred candidates for data efficiency. 
These selected structures were computed with DFT, added to the original training set, and used to upfit the ACE potential. 
As a result of the above approach, we observed a significant decrease in the extrapolation grades for representative nanocluster (see Supplementary Fig. S3), and a reduction in the error metrics for the energy and forces of the selected surface structures. 
The energy root mean square error (RMSE) decreased from 31 to 3 meV/atom and the force RMSE from 172 to 84 meV/\AA (see Supplementary Fig. S1).

\subsection{STEM characterization}
A probe corrected Titan Themis 60-300 operated at 300 kV was used for the STEM characterization. For imaging, a HAADF detector with a collection angle of 78-200 mrad was used. The EDS composition maps were acquired with a Brucker Super X detector. 

\backmatter

\bmhead{Supplementary information}

The supplementary information is provided in the following PDF file.

\bmhead{Acknowledgments}
The authors acknowledge computation time by Center for Interface-Dominated High Performance Materials (ZGH) at Ruhr-Universit{\"a}t Bochum, Germany.

\bmhead{Funding} 
This work was in part supported by the German Science Foundation (DFG), projects 405621081 and 405621217.

\bmhead{Data availability}
DFT reference data that was used for training and the ACE potential file are available upon request.

\bmhead{Conflict of interest}
On behalf of all authors, the corresponding author states that there is no conflict of interest.

\bibliography{PtRh}% common bib file

\end{document}

\title[Supplementary information]{Supplementary Information\vspace{0.5cm}\\ 
\large {Atomic cluster expansion for Pt-Rh catalysts: From ab initio to the simulation of nanoclusters in few steps}}

%%=============================================================%%

\author[1]{\fnm{Yanyan} \sur{Liang}}\email{yanyan.liang@rub.de}

\author[1]{\fnm{Matous} \sur{Mrovec}}\email{matous.mrovec@rub.de}

\author[1]{\fnm{Yury} \sur{Lysogorskiy}}\email{yury.lysogorskiy@rub.de}

\author[2]{\fnm{Miquel} \sur{Vega-Paredes}}\email{m.vega@mpie.de}

\author[2]{\fnm{Christina} \sur{Scheu}}\email{scheu@mpie.de}

\author*[1]{\fnm{Ralf} \sur{Drautz}}\email{ralf.drautz@rub.de}

\affil*[1]{\orgdiv{ICAMS}, \orgname{Ruhr-Universit{\"a}t Bochum}, \orgaddress{\street{Universit{\"a}tsstra{\ss}e 150}, \city{Bochum}, \postcode{44801}, \country{Germany}}}

\affil[2]{\orgname{Max Planck Institut für Eisenforschung}, \orgaddress{\street{Max Planck Straße 1}, \city{Düsseldorf}, \postcode{40237}, \country{Germany}}}

\maketitle

\begin{figure}
    \centering
    \includegraphics[scale=0.5]{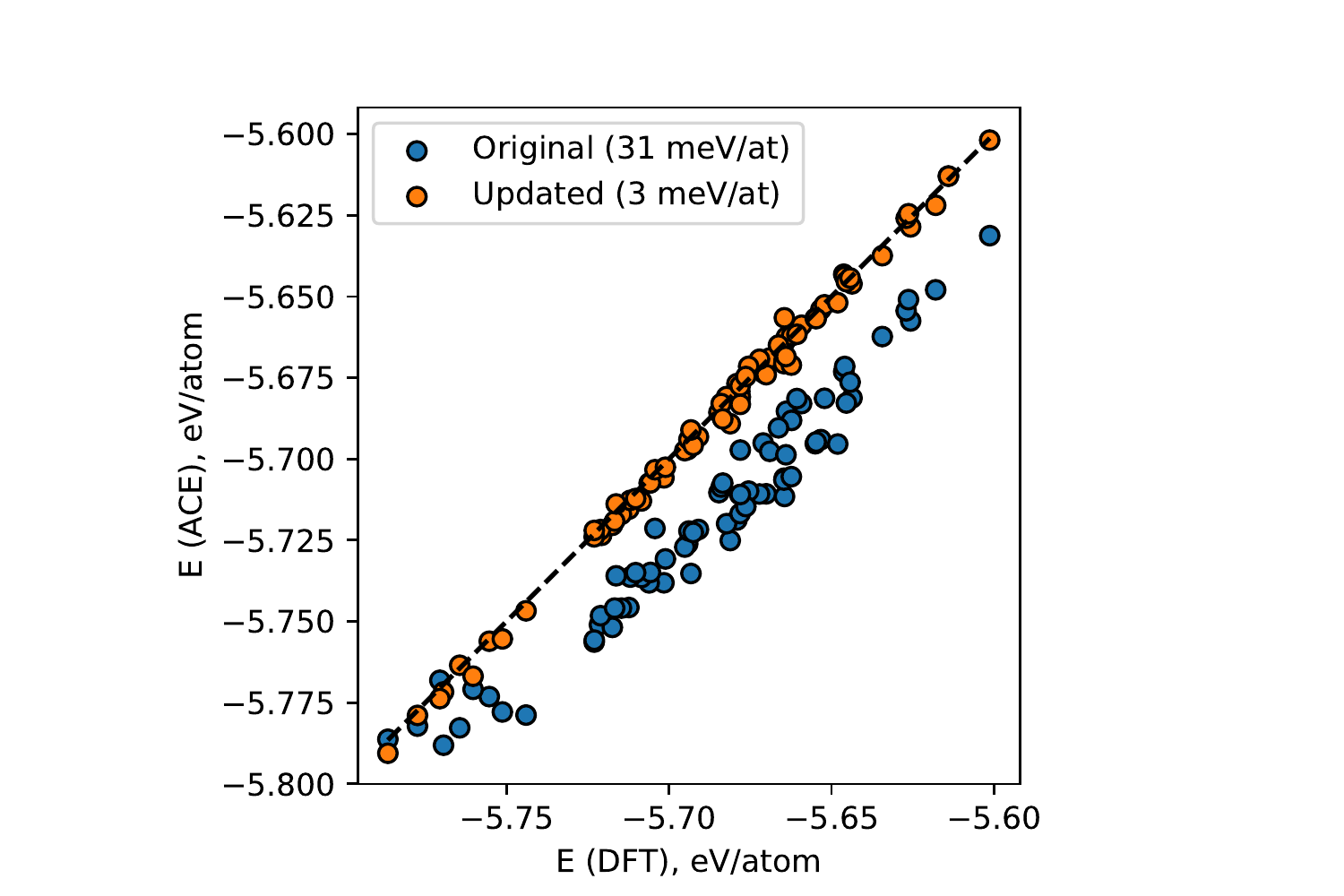}
    \includegraphics[scale=0.5]{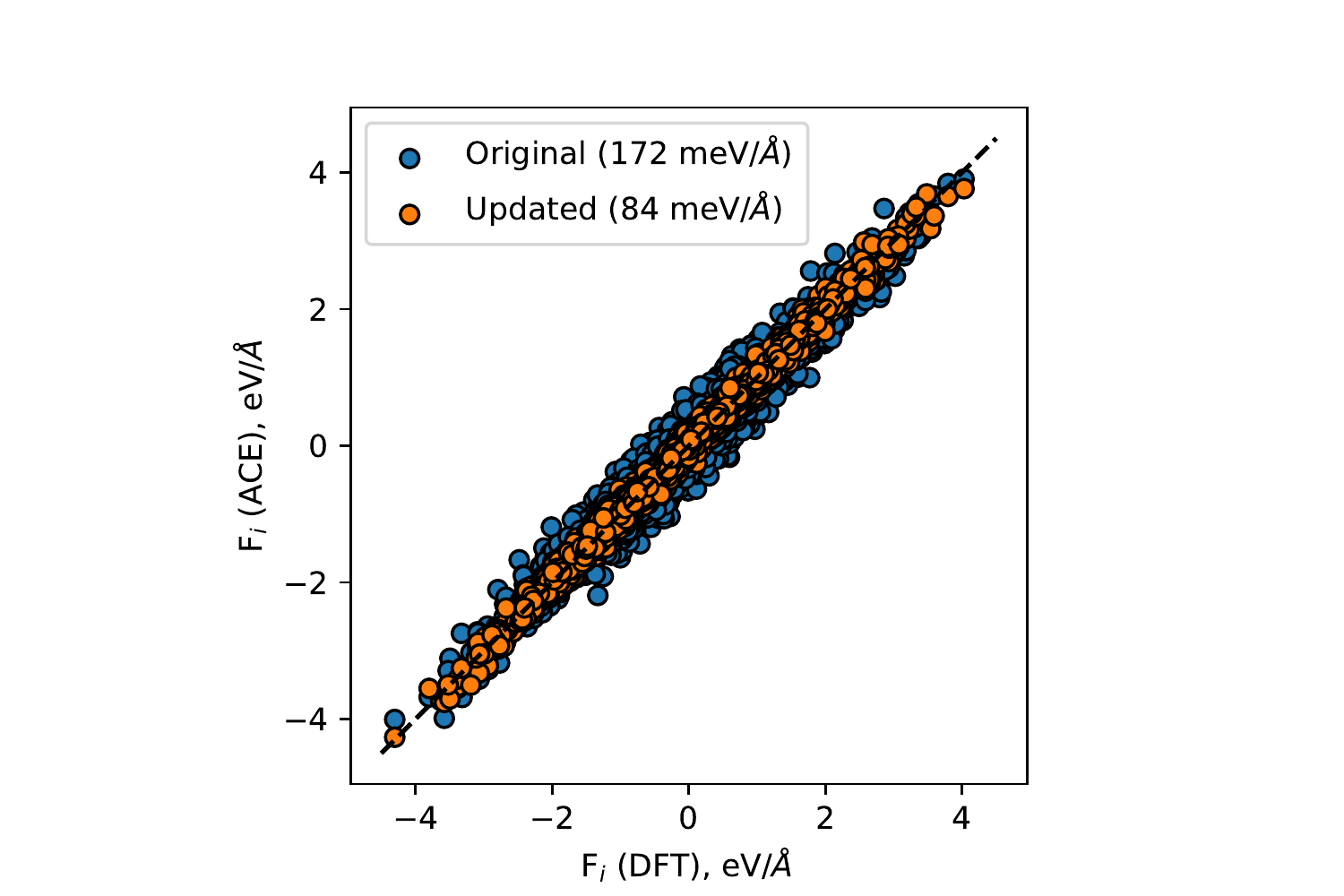}
    \caption{Energy (top figure) and forces (bottom figure) error metrics on selected surface configurations before and after upfitting of ACE potential.}
    \label{fig:AL_config_metrics}
\end{figure}

\begin{figure}[htbp]
    \centering
    \includegraphics[scale=0.3]{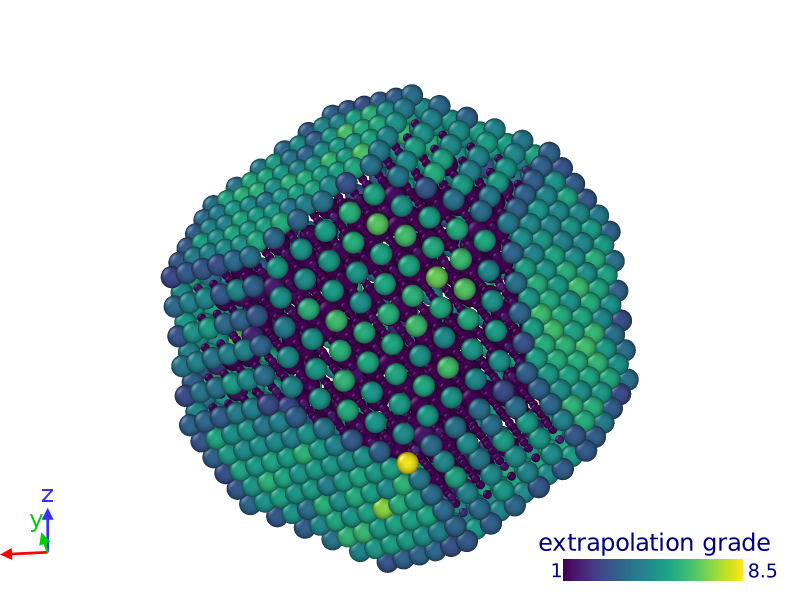}
    \includegraphics[scale=0.3]{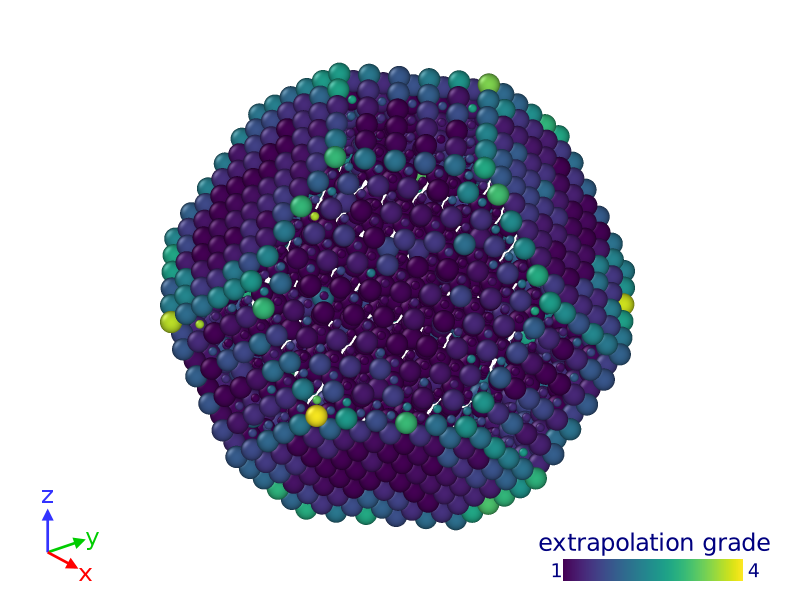}
    \caption{A comparison showing the extrapolation grades for the surface atoms of the octahedron clusters. The top configuration was evaluated using the initial version of the ACE parametrization, the bottom configuration was simulated using the retrained ACE parametrization after the active learning procedure. }
    \label{fig:gamma_cluster}
\end{figure}

\begin{figure}[htpb]
    \centering
    \subfigure[$\text{Pt}_{30}\text{Rh}_{70}$]{\includegraphics[width=0.42\textwidth]{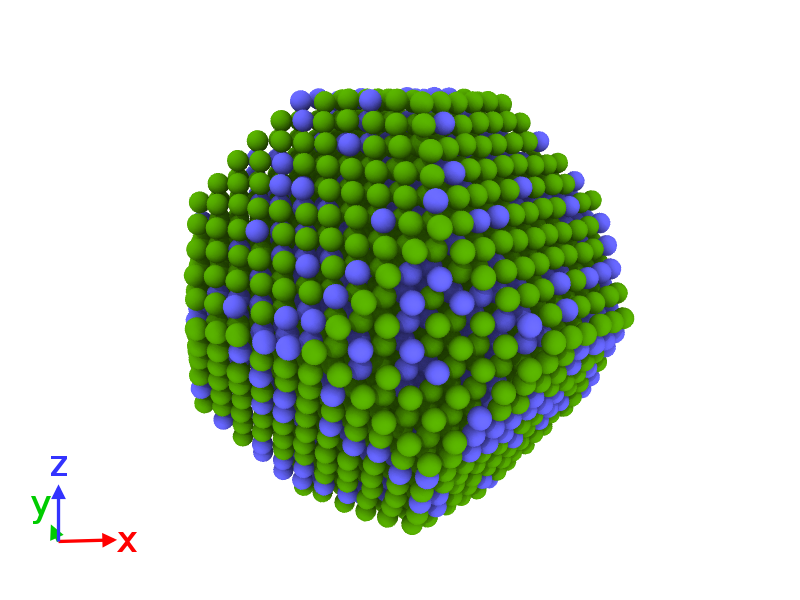}}
    \subfigure[[$\text{Pt}_{45}\text{Rh}_{55}$]{\includegraphics[width=0.42\textwidth]{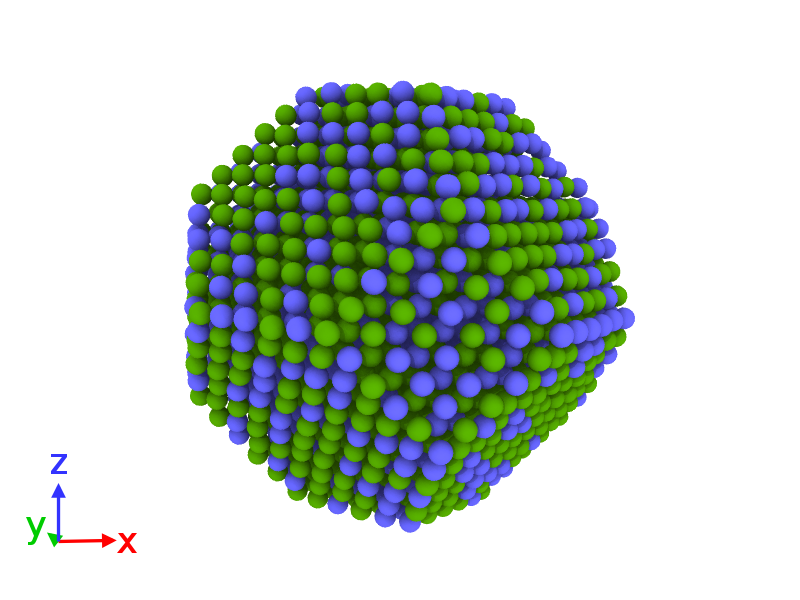}}
    %\subfigure[]{\includegraphics[width=0.32\textwidth]{image/Pt30Rh70_final_cut.png}}
    \caption{Visualization of the initial configurations of the octahedron nanoclusters used in the hybrid MD/MC simulations. Pt (blue) and Rh (green) atoms are randomly distributed in the nanoclusters. 
    }
    \label{supp-fig:twonano}
\end{figure}

\begin{figure}
    \centering
    \includegraphics[scale=0.50]{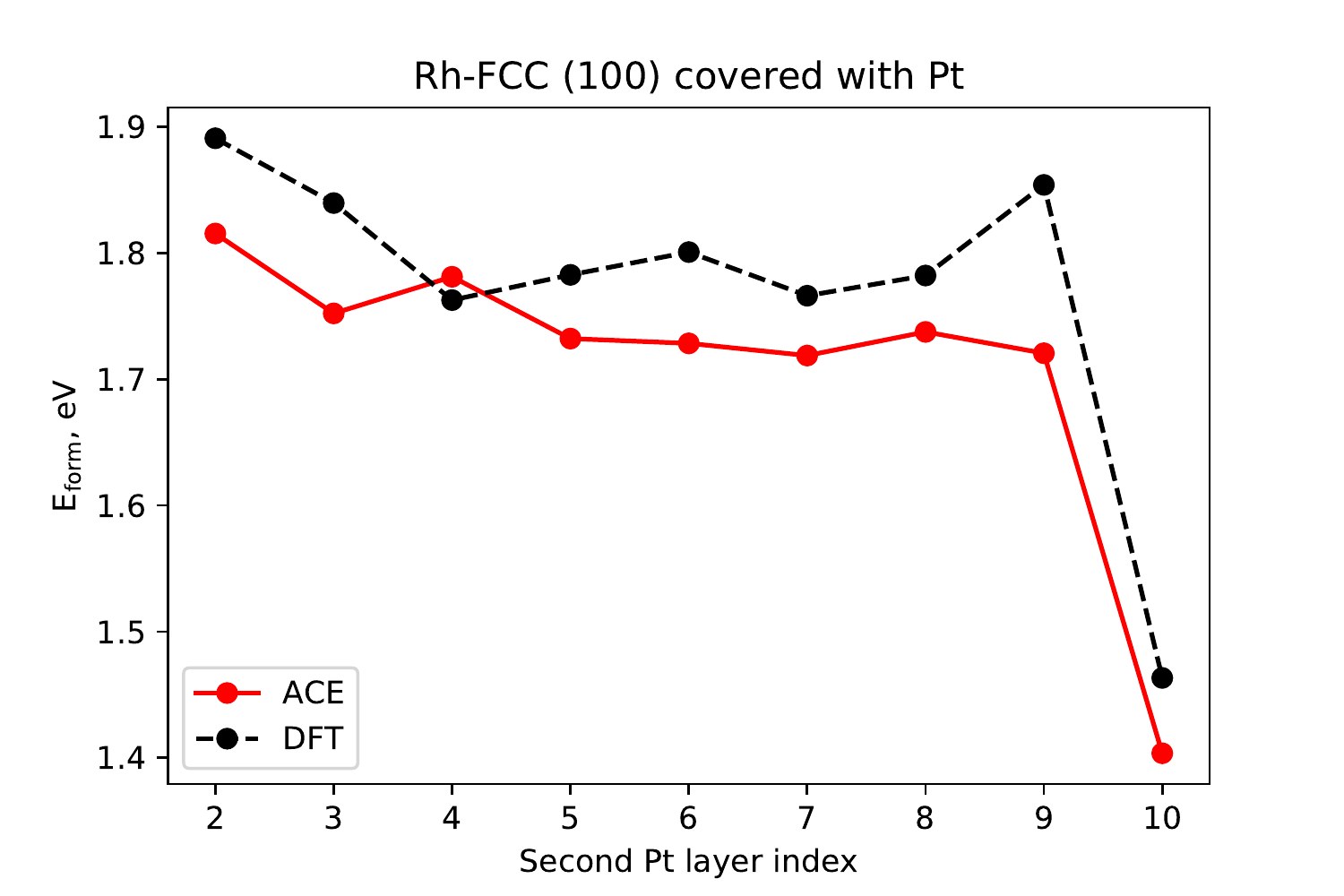}

    \includegraphics[scale=0.50]{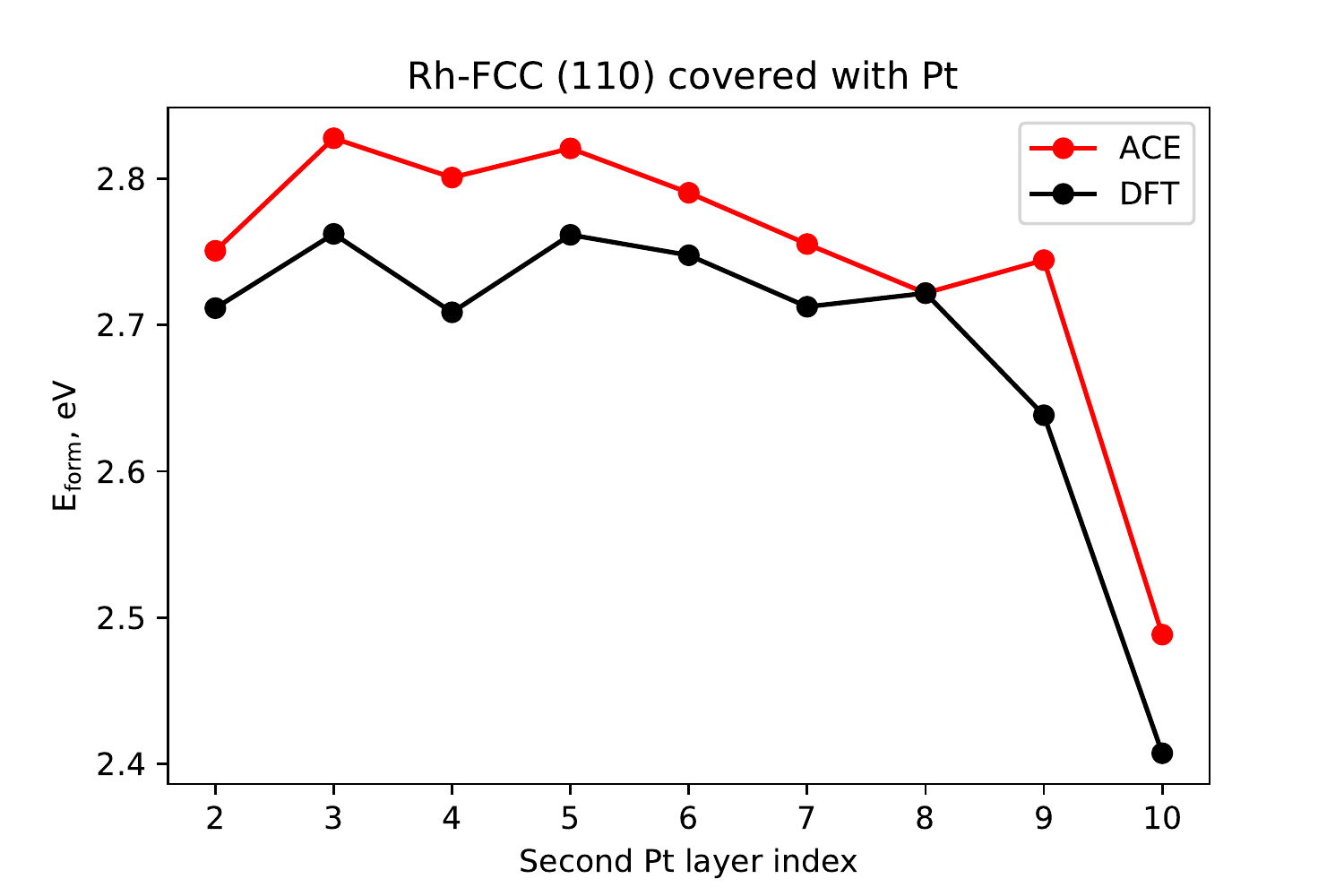}

    \includegraphics[scale=0.50]{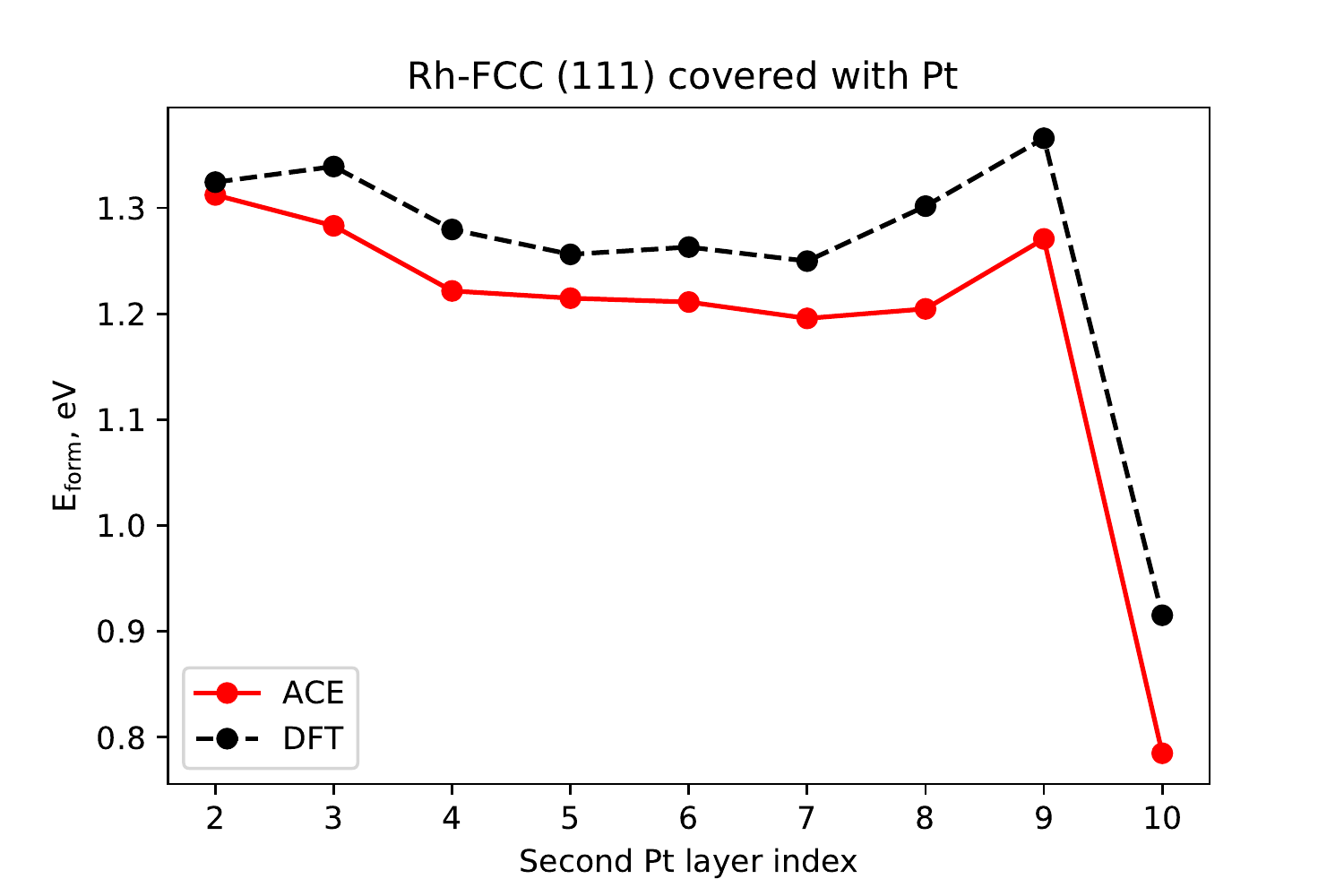}
    \caption{
    Dependence of formation energies of surface slabs consisting of two Pt and eight Rh atomic layers. The first Pt layer is always positioned on the surface (index 1) while the position of the second Pt layer varies (i.e., index 2 for the 2nd Pt layer corresponds to two adjacent Pt layers at one surface of the slab while index 10 corresponds to a Rh slab covered on both surfaces with Pt).  The formation energies are computed with respect to energies of the ground state fcc Pt and Rh. 
    }
    \label{fig:rh_coseg}
\end{figure}